\documentclass[a4paper,fleqn,usenatbib]{mnras}
\pdfoutput=1

\usepackage{txfonts}

\usepackage[T1]{fontenc}

\usepackage{graphicx}
\usepackage{graphics}
\usepackage{xspace}
\usepackage{tabularx}
\newcolumntype{Y}{>{\centering\arraybackslash}X}  

\usepackage{amssymb}    
\usepackage{color}

\renewcommand{\d}{\mathrm{d}}

\newcommand{\ie}{{i.e.~}}
\newcommand{\eg}{{e.g.~}}

\newcommand{\LCDM}{$\Lambda$CDM\xspace}

\newcommand{\pgadget}{{\sc PGADGET-3}\xspace}
\newcommand{\music}{{\sc MUSIC}\xspace}
\newcommand{\AHF}{{\sc AHF}\xspace}
\newcommand{\virg}[1]{`{#1}'}
\newcommand{\citenp}[1]{\citeauthor{#1} \citeyear{#1}}  

\newcommand{\Rh}{R_{\rm h}}
\newcommand{\Rg}{R_{\rm g}}

\newcommand{\Mh}{M_{\rm h}}
\newcommand{\Msh}{M_{\rm sh}}

\newcommand{\Vmax}{V_{\rm max}}
\newcommand{\hMsol}{h^{-1}\,{\rm M_\odot}}
\newcommand{\hMpc}{h^{-1}\,{\rm Mpc}}
\newcommand{\hkpc}{h^{-1}\,{\rm kpc}}
\newcommand{\zc}{z_{\rm c}}
\newcommand{\tc}{t_{\rm c}}
\newcommand{\ac}{a_{\rm c}}
\newcommand{\vr}{\varv_{\rm r}}
\newcommand{\rs}{r_{\rm s}}
\newcommand{\fsub}{f_{\rm sh}}
\newcommand{\Nsub}{N_{\rm sh}}

\newcommand{\Rspl}{R_{\rm spl}}
\newcommand{\fdisc}{f_{\rm disc}}

\newcommand{\zac}{z_{\rm ac}}
\newcommand{\tac}{t_{\rm ac}}
\newcommand{\Mac}{M_{\rm ac}}
\newcommand{\Lcluster}{$L$-cluster\xspace}
\newcommand{\Lclusters}{$L$-clusters\xspace}

\newcommand{\paperI}{paper I\xspace}
\newcommand{\paperII}{paper II\xspace}

\newcommand{\amun}{Amun\xspace}   
\newcommand{\abu}{Abu\xspace}     
\newcommand{\supay}{Supay\xspace} 
\newcommand{\siris}{Siris\xspace} 

\newcommand{\old}{stalled\xspace}
\newcommand{\young}{accreting\xspace}
\newcommand{\Old}{Stalled\xspace}

\title[ZOMG III: The satellite population]
{ZOMG III: The effect of Halo Assembly on the Satellite Population}

\author[E. Garaldi et al.]{
Enrico Garaldi,\thanks{egaraldi@uni-bonn.de}\thanks{Member of the International Max Planck Research School (IMPRS) for Astronomy and Astrophysics at the Universities of Bonn and Cologne}
Emilio Romano-D\'{\i}az,
Mikolaj Borzyszkowski and Cristiano Porciani
\\
Argelander Institut f\"ur Astronomie, Auf dem H\"ugel 71, Bonn, D-53121, Germany}

\date{Accepted XXX. Received YYY; in original form ZZZ}

\pubyear{2017}

\begin{document}
\label{firstpage}
\pagerange{\pageref{firstpage}--\pageref{lastpage}}
\maketitle


\begin{abstract}
We use zoom hydrodynamical simulations to investigate the properties of satellites within galaxy-sized dark-matter haloes with different assembly histories. We consider two classes of haloes at redshift $z=0$: \virg{\old} haloes that assembled at $z>1$ and \virg{\young} ones that are still forming nowadays. Previously, we showed that the \old haloes are embedded within thick filaments of the cosmic web while the \young ones lie where multiple thin filaments converge.
We find that satellites in the two classes have both similar and different properties.
Their mass spectra, radial count profiles, baryonic and stellar content, and the amount of material they shed are indistinguishable.
However, the mass fraction locked in satellites is substantially larger for the \young haloes as they experience more mergers at late times. The largest difference is found in the satellite kinematics. Substructures fall towards the \young haloes along quasi-radial trajectories whereas an important tangential velocity component is developed,
before accretion, 
while orbiting the filament that surrounds the \old haloes.
Thus, the velocity anisotropy parameter of the satellites ($\beta$) is positive for the \young haloes and negative for the \old ones. This signature enables us to tentatively categorize the Milky Way halo as \old based on a recent measurement of $\beta$.
Half of our haloes contain clusters of satellites with aligned orbital angular momenta corresponding to flattened structures in space. These features are not driven
by baryonic physics
and are only found in haloes hosting grand-design spiral galaxies, independently of their assembly history.
\end{abstract}



\section{Introduction}
\label{sec:intro}
The standard model of cosmology is based on conventional physics and assumes that
the dominant terms of the stress-energy tensor are a cosmological constant and cold dark matter. In this $\Lambda$CDM scenario, 
self-gravitating structures form hierarchically through mergers of smaller units.
The end product of this process is a collection of dark matter haloes containing
a large amount of substructures that are the vestiges of the merging process.
In fact, these `satellites' enter, orbit and get progressively stripped of their outer layers while inside the host halo.

It was originally thought that substructures would be quickly erased but
the advent of high-resolution $N$-body simulations revealed that this is not the case \citep[e.g.][]{Moore+1996, Tormen+1998, Klypin+1999}. Comparing the computer models 
with the population of Milky-Way satellites posed several challenges
to the \LCDM model \citep[see e.g.][for a review]{Kravtsov2010, Bullock2010, Weinberg+2015}.
Recent studies have characterized the detailed statistical properties of the surviving satellites \citep{Gao+2011, Ishiyama+2013, Wu+2013} and shown that they are very sensitive to gas and star-formation
physics \citep[\eg][]{Dolag+2009, Okamoto+2010, Romano-Diaz+2010, Schewtschenko+2011, DiCintio+2013, Wang+2016, Chua+2016}.

This article is the last in a series of three introducing a numerical project 
named Zooming On a Mob of Galaxies (ZOMG). 
ZOMG uses a suite of zoom $N$-body plus hydrodynamical simulations
to study how the cosmic environment regulates the evolution and properties of galaxy-sized dark-matter haloes as well as of their baryonic content and substructures.
It is well known that haloes of the same mass show 
different clustering properties depending on their formation history,
a process usually dubbed \virg{assembly bias} \citep{Gao+2005,Harker+2006, Zhu+2006, Gao+2007, Dalal+2008}. A long-standing challenge in theoretical cosmology has been to understand
the origin of this phenomenon, especially for galaxy-sized haloes.
Building upon the early work by \citet{Hahn+2009}, in \citet[][hereafter \paperI]{paperI}, we have shown that haloes stop growing in mass once they are embedded in prominent filaments of the cosmic web that
are thicker than the halo diameter.
The dark matter (DM) in these haloes preferentially follows tangential orbits due to the gravitational
pull of the filament that alters the trajectories of the infalling material before it reaches the halo. 
Conversely, haloes sitting at the knots of the cosmic web (the regions towards which numerous thin filaments converge) grow by accreting material from the surroundings along quasi-radial orbits.
The accretion history and the internal dynamics of galaxy-sized haloes are thus intimately
related to the halo location within the cosmic web, hence the assembly bias. 
It is natural to ask whether the fate of the gas component also depend on the halo
assembly history and position. In \citet[][hereafter \paperII]{paperII}, we have addressed this question 
showing that the properties of the central galaxy are largely insensitive to the collapse time of the host, with the exception of the thickness and age of its stellar disc (both increasing for haloes with stalled growth in filaments).
Finally, 
in this work, we investigate the impact of the halo assembly history (and thus of cosmic environment) on the properties of the satellite population.

Several related lines of research have been recently pursued using numerical simulations.
Although they have identified the existence of deep interrelationships between halo environment,
matter accretion and the final characteristics of the substructures, a clear picture has not yet emerged.
The connection between cosmic filaments and the kinematics of substructures has been particularly explored in the literature. 
Infall along filaments seem to produce groups of subhaloes whose orbital angular momenta align with the halo spin in both possible rotating directions \citep{Lovell+2011}.
Even the survival lifetime of substructures seem to depend on whether they have been accreted along
a filament or not \citep{Gonzalez+Padilla-2016}. 
It has been concluded that the ordered
accretion of substructures along filaments is the prime reason for the existence of flattened configurations of satellite galaxies which are coherently rotating \citep{Libeskind+2005,Libeskind+2015,Lovell+2011}.
Some authors, however, argue that such layouts can only exist when at most two filaments feed the host halo \citep{Ahmed+2017}.
Others find that thin planes of satellites are only hosted by haloes with very concentrated 
mass density profiles which form early on when filaments are narrower and accretion is thus
more focused \citep{Buck+2015}.
In general, the spatial distribution of the satellites seem to align (to some extent) with the large-scale distribution of matter surrounding the host halo and even with the halo shape \citep{Shao+2016}.
In fact, substructures preferentially fall in along the main principal axis of the inertia tensor \citep{Libeskind+2014}.

In this paper, we use the high-resolution simulations presented in \paperI and \paperII to
study how the population of satellite galaxies is influenced by the halo assembly history.
The article is structured as follows. In Section \ref{sec:methods}, 
we describe the main features of the ZOMG simulations and the analysis performed for this work. The detailed time evolution of a few substructures is described in Section \ref{sec:sat_evolution} while, in Section \ref{sec:Comparison}, we study the statistical properties of the satellite populations
providing comparisons with previous numerical studies and observations.
In Section \ref{sec:assembly}, we tackle the issue of how the process of halo assembly
impacts the kinematics and the final spatial configuration of the satellites.
A summary of our results and conclusions are presented 
in Section \ref{sec:conclusions}.


\section{Numerical Methods}
\label{sec:methods}

\begin{table*}
  \caption{Properties of the re-simulated haloes at $z=0$. 
 From left to right, the columns give: the name, the mass ($\Mh$), the radius ($\Rh$), the scale radius ($\rs$) obtained fitting the radial mass-density profile with the Navarro-Frenk-White (NFW) formula
\citep{NFW}, 
the expansion factor at collapse time ($\ac$), the total number of substructures ($\Nsub$), the number of substructures with gas ($N_{\rm gas}$), the number of substructures with stars ($N_{*}$), the fraction of $\Mh$ in substructures ($\fsub$), the mean fraction of the baryonic content of substructures identified at $z=2$ that migrates to the central galaxy by $z=0$ ($\langle \fdisc \rangle$) and the colour associated to the halo throughout the paper.}
\begin{tabularx}{\textwidth}{XYYYYYYYYYYY}
\hline
Halo & $\Mh$ & $\Rh$ & $\rs$ & $\ac$ & $\Nsub$ & $N_{\rm gas}$ & $N_{*}$ & $\fsub$ & $\langle \fdisc\rangle$ & colour \\
name & [$10^{11}\,\hMsol$] & [$\hkpc$] & [$\hkpc$] & & & & & & & \\

\hline
\abu    &  $4.1$  &  151  &  3.13  &  0.968  &  382  &  4  &  52  &  0.109  &  0.03 &  \includegraphics{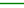}\\
\amun   &  $3.5$  &  144  &  7.41  &  >1.00  &  477  &  4  &  73  &  0.100  &  0.05 &  \includegraphics{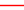}\\
\supay  &  $4.3$  &  153  &  5.05  &  0.404  &  647  &  3  &  97  &  0.067  &  0.09 &  \includegraphics{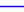}\\
\siris  &  $2.3$  &  124  &  3.02  &  0.333  &  228  &  1  &  32  &  0.067  &  0.03 &  \includegraphics{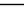}\\
\hline

\end{tabularx}
\label{table:Halo}
\end{table*}

We summarize here the main properties of the ZOMG simulations and
discuss the substructure analysis which forms the main focus of the paper.
Further details can be found in papers I and II. 

\begin{figure*}
\begin{center}
\includegraphics[height=0.9\textheight]{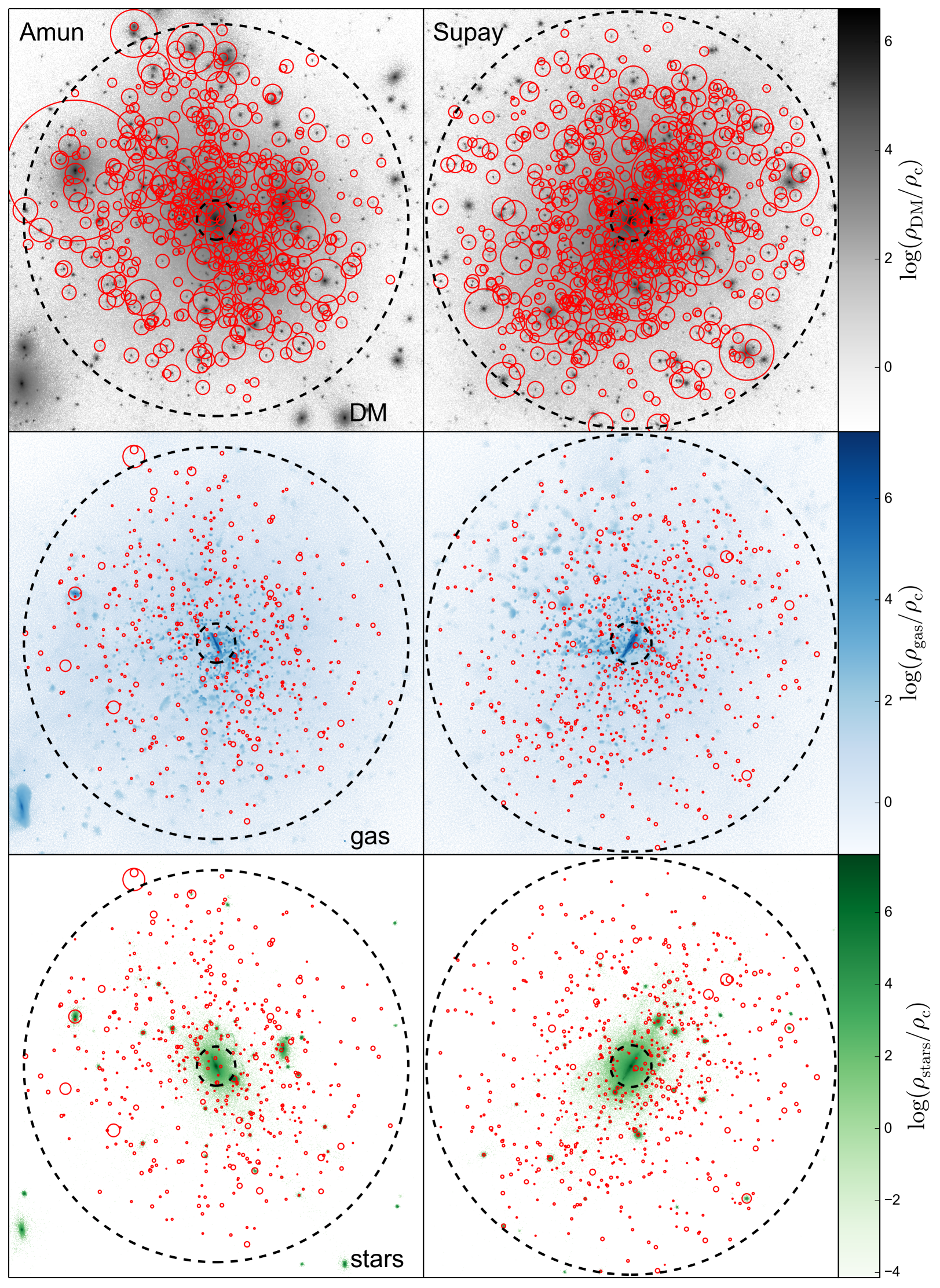}
\end{center}
\caption{Dark-matter, gas and stellar distributions (from top to bottom) for two of our re-simulated haloes at $z=0$.
Each panel shows the projection along one axis of the simulation particles contained within a cube of side $310 \, \hkpc$ centred on the corresponding halo. Particles are colour-coded according to the local density computed using a standard SPH-like kernel with 64 neighbours (of the same species), normalized by the critical density of the universe $\rho_{\rm c}$.
The outer and inner dashed circles indicate the halo radius $\Rh$ and the radius of the central
galaxy $\Rg$, respectively. Substructures are highlighted with solid red circles whose radius
reflects the subhalo radius found by \AHF (top panels) or the location of the maximum circular velocity
(middle and bottom panels). Note that dense DM clumps that are not surrounded by a red circle
are located outside the halo radius although they might give the impression to be within $\Rh$ due to projection effects.
}
\label{fig:finder}
\end{figure*}

\label{sec:classification}
The ZOMG project includes a set of high-resolution $N$-body and hydrodynamical simulations 
that follow structure formation in a model universe with
$\Omega_m = 0.308$, $\Omega_{\Lambda} = 0.692$, $\Omega_b = 0.0481$ and
$h = 0.678$ \citep{Planck2013_Cosmology}. Linear density perturbations are characterized by
the spectral index $n = 0.9608$ and the power-spectrum normalization $\sigma_8 = 0.826$.
All simulations 
cover the same
periodic cubic box with a side of $50 \, \hMpc$.
Initial conditions are generated at redshift $z=99$ using the \music code \citep{MUSIC} and 
employing second-order Lagrangian perturbation theory to shift particles from a uniform Cartesian grid.

To isolate the host haloes of present-day $L_*$ galaxies,
we make use of a parent $N$-body run containing $512^3$ identical particles and select a few objects with masses $\Mh \sim {\rm few} \times 10^{11} \, \hMsol$ at $z=0$ that we then re-simulate at very high-resolution using the multimass zoom technique.
On top of the mass selection, we apply a further criterion based on  
the assembly history of the haloes that we characterize in terms of the `collapse time' introduced in \citet{Borzyszkowski+2014}.
In brief, we trace the particles that form a halo at $z=0$ back in time
and compute the evolution of their tensor of inertia.
Given the inertia ellipsoid, we rigidly rescale it so that it always contains the final
mass of the halo and we calculate its volume $V(t)$.
This procedure was conceived to follow the evolution of the
outermost matter shell forming the halo at $z=0$, the collapse of which ultimately 
determines the halo assembly time. 
The function $V(t)$ initially increases due to the Hubble expansion, reaches a maximum (that can be used to define the epoch of turnaround) and generally decreases afterwards until it starts oscillating around
a constant value indicating virialization (actual examples are shown in Fig. 1 of \paperI).
The collapse time of a halo, $\tc$ (or the corresponding redshift $\zc$ and expansion factor $\ac$), 
is defined as the moment at which the volume stabilizes (see \paperI for details regarding the practical
implementation of this calculation).
Haloes with fixed mass in the parent run show a broad range of collapse times. 
To maximize the differences among the re-simulated haloes and
better recognize the effects of their assembly histories, we only pick objects in the tails of the distribution.
For the $N$-body (DM-only) runs (see \paperI), we thus randomly pick 5 
\virg{\old} ($\zc > 1$) and 2 \virg{\young} ($\zc \lesssim 0$) objects
among the haloes of the selected mass identified in the parent simulation at $z=0$.
For the much more time consuming hydrodynamical simulations (see \paperII), instead, we only consider 2 haloes for each class.
We adopt the nomenclature of papers I and II where each re-simulated halo is named 
after an ancient god, sharing the initial letter of its name with the 
class it belongs to (\ie \virg{S} for \old and \virg{A} for \young haloes). 

The re-simulations are carried out using a modified version of the
tree-particle-mesh smoothed particle hydrodynamics (SPH) code \pgadget
\citep{gadget2}. Our hydrodynamic runs include radiative cooling, star
formation and stellar feedback, galactic winds, a multi-phase
interstellar medium \citep{springel-hernquist+03} and an ultraviolet
background active from $z \sim 11$ that reionizes hydrogen in the intergalactic medium by $z \sim 6$ \citep{Haardt+2001}.
In all cases,
we achieve an effective resolution of $4096^3$ computational elements in the region of interest
(roughly extending up to three times the halo radius), corresponding to particle masses of
$m_{\rm DM} = 1.31 \times 10^5 \, \hMsol$ and $m_{\rm gas} = 2.43 \times
10^4 \, \hMsol$ for DM and gas, respectively. Each gas particle 
can experience up to two episodes of star formation, in each of which a mass $m_{\rm gas}/2$ is converted
into a stellar particle.
Our runs extend to $z=0$. We save a few snapshots at early times plus one every 20 Myr after redshift
$z=9$, for a total of 682 output files.

\label{sec:identification}
We identify gravitationally bound objects and their substructures  
using the {\sc Amiga Halo Finder} \citep[\AHF,][]{AHF1,AHF2}.
This software initially defines haloes as spherical regions
with a mean matter density of $200$ times 
the critical density of the universe, $\rho_{\rm c}(z)$,
and then iteratively purges them of the (unbound) particles that
move faster than 1.5 times
the escape velocity.
The halo radius, $\Rh$, and mass, $\Mh$, are defined using the smallest sphere enclosing all the bound
particles.
We conventionally identify the `central galaxy' of each halo with the innermost region of
radius $\Rg = 0.1\, \Rh$ \citep[\eg][]{Scannapieco+2012}.
Following \citet{Diemer+2014}, we also consider a second definition for the halo boundary 
by locating a sharp steepening of the radial mass-density profile.
\citet{Adhikari+2014} and \citet{More+2015} argued that 
such radius correspond to the first apocentre of recently accreted matter 
and therefore dubbed it as the \virg{splashback radius}, $\Rspl$. Contrary to the halo radius, 
$\Rspl$ is parameter free and does not suffer from pseudo evolution due to change of $\rho_{\rm c}$ with time.

We use the built-in function of \AHF to construct halo merger trees. 
The main progenitor of a halo is selected maximizing 
the merit function $N^2_{i \cap j} / N_i N_j$, 
where $N_i$ and $N_j$ denote the number of particles in the progenitor and the descendant
in two consecutive snapshots
and $N_{i \cap j}$ is the number of particles they share.

\AHF automatically detects 
substructures as density peaks within a main halo. Their edge
is initially determined as the minimum of the radial
density profile and then adjusted to the radius of the smallest sphere
enclosing all bound particles.
Note that we
do not consider higher levels of (nested) substructures (\ie sub-subhaloes are not distinguished from their host subhaloes).
Substructures are dynamic entities that continuously accrete and loose material.
When studying their evolution, we follow all the particles they are made of at the time they reach
their maximum mass (typically, right before entering the host).

\label{sec:sh_analysis}

Fig. \ref{fig:finder} gives a visual impression of two of our re-simulated haloes 
and of their substructures.
The main properties of the four haloes analyzed in this paper are listed in Table \ref{table:Halo}.
These data are extracted from the hydrodynamic simulations, corresponding results for 
the $N$-body runs can be found in Table 1 of \paperI.

\section{Satellite evolution}
\label{sec:sat_evolution}

\begin{figure*}
\begin{center}
\includegraphics[width=0.95\textwidth]{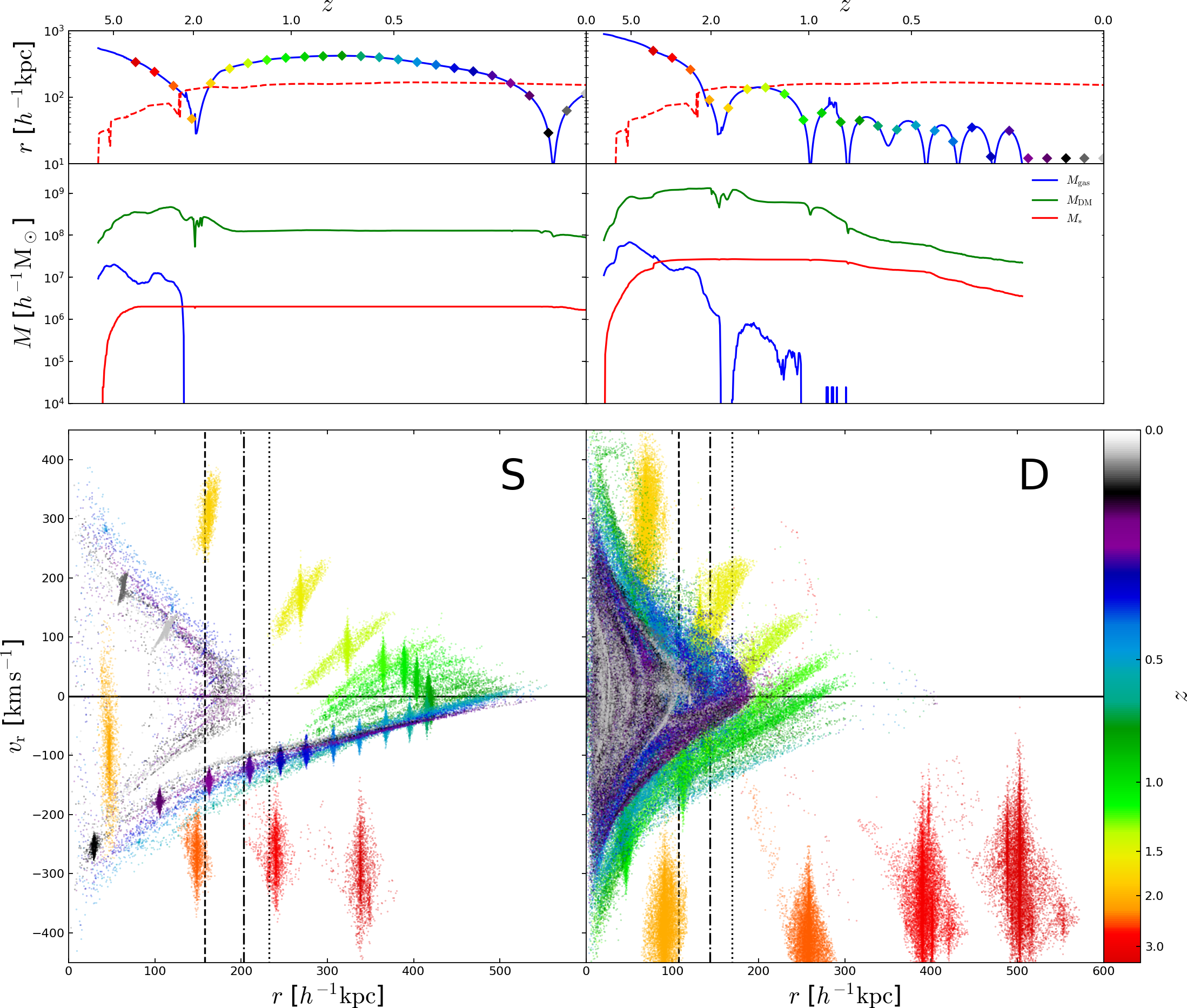}
\end{center}
\caption{Evolution of different properties in two of the
  most massive subhaloes in our simulation suite (the left one belongs to \supay 
  while the right one is part of \amun).
  Top: evolution of the distance $r$ from the host. The dashed red line denotes the 
  host radius. The diamond symbols indicate the redshift of the steps
  plotted in the bottom panels.
  Middle: evolution of the DM, gas and stellar mass of the substructure 
  (green, blue and red line, respectively).
  Bottom: radial phase space distribution of all particles. 
  The colour encodes the redshift, in steps of 
  $500 \, \rm{Myr}$. The substructure in the left panel has survived 
  until $z=0$ while the one in the right panel has been totally disrupted. 
  The halo radius, the splashback radius 
  and its predicted (median) value for haloes with the same mass and accretion rate 
  \citep[following][]{More+2015} are shown at the redshift of the first 
  apocentre using dashed, dot-dashed and dotted lines respectively.}
\label{fig:phase_space_evolution}
\end{figure*}

\begin{figure*}
\begin{center}
\includegraphics[width=0.98\textwidth]{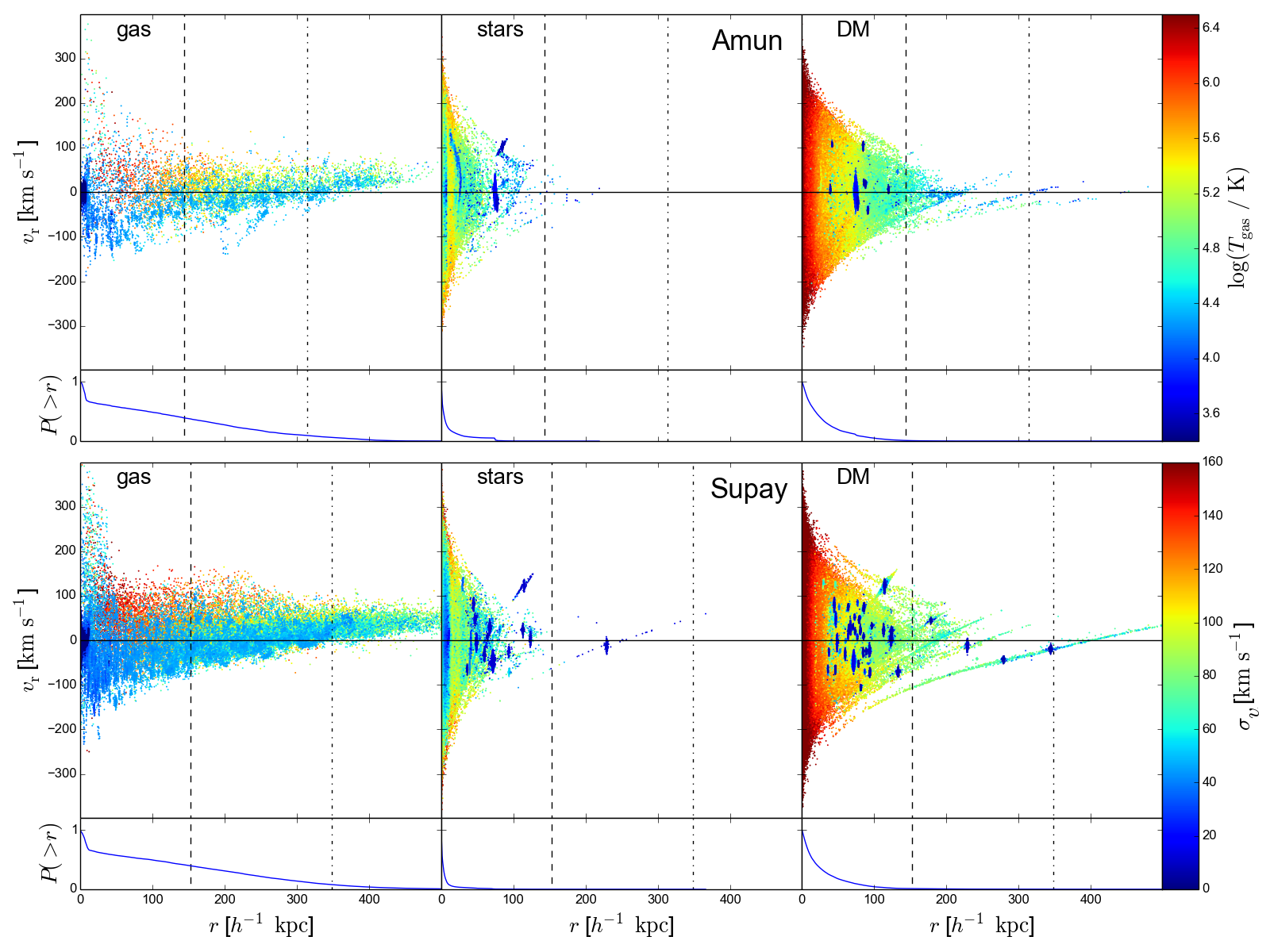}
\end{center}
\caption{Radial phase space distribution at $z=0$ of particles of different species
  belonging to the 100 most massive subhaloes identified at $z=2$ in
  \amun (top) and \supay (bottom). The particles are
  colour-coded according to the temperature of the gas (left) and the
  velocity dispersion of stars (middle) and DM (right).
  The inset on the bottom of each panel shows the
  cumulative probability distributions of finding a particle at a
  distance \textit{larger} than a given radius.
Note that the cold gas with high (positive) radial velocities at very small
radii is due to the numerical treatment of supernova feedback in the central galaxy.
  }
\label{fig:phase_space_z0}
\end{figure*}

In this section, we exploit the excellent time- and mass-resolution of the ZOMG simulations to describe the evolutionary path of a few typical substructures. This study helps the reader recognize the complex phenomenology of
satellite-host interactions
and us introduce several key concepts that will be used in the remainder of the paper.
Ultimately, we determine the amount of material that satellites shed to their host halo and its central galaxy.

\subsection{Trajectories and dynamics}
We conventionally define the accretion redshift of a satellites, $\zac$, by identifying the instant when, for the first time, 
AHF associates it with a host halo.
The subsequent fate of the satellite is regulated by the interplay between several physical
mechanisms \citep[e.g.][and references therein]{Binney_Tremaine_book, Mo_vdB_White_book, Mayer+2010}. 
Gravitational interactions between the satellite and the diffuse material that makes up the host produce a net drag known as dynamical friction.
As a result, the satellite looses energy and angular momentum and its orbit
decays towards the central region of the host where the gravitational potential
reaches its minimum value.
Along the way, the satellite constantly looses matter from its outer parts due to
the action of tidal forces (tidal stripping). The ejected material initially 
forms leading and trailing streams stretching for large distances compared
with the core of the satellite. Later on, these tidal tails evolve into
approximately spherical shells and eventually phase-mix with the  
diffuse component of the host halo.
In the CDM scenario, most satellites accrete on to their hosts following highly eccentric
orbits and thus experience time-varying tidal forces.
At each pericentric passage, when tides become particularly strong, the satellite expands, gains kinetic energy and rearranges its internal structure  
(tidal heating). Its lower binding energy makes it prone to further mass loss via tidal stripping (and ram pressure stripping for the gas component).
Since the orbital decay rate depends on the satellite mass, all these effects are
highly interconnected. The detailed evolution thus depends on the inital mass and concentration as well as the orbital parameters of the satellite.

In order to provide a few representative examples,
in Fig. \ref{fig:phase_space_evolution} we consider two rather massive substructures extracted from our simulations.
Although they both have $\zac \approx 2$,
one of them survives till the present time while the other is completely disrupted (meaning that \AHF cannot identify it any longer at late times against the background of the host halo).
For this reason,
we dub them S-Sat and D-Sat, respectively. S-Sat (shown on the left-hand side) is part of Supay while D-Sat (displayed on the right-hand side) is hosted by Amun, although this is not important as similar examples are present in every re-simulated halo. 
In the top panels, we show the distance of the satellites from the centres of the corresponding hosts as a function of time. The curves start well before the satellites first `enter' their host and follow their radial trajectories until the present time.
S-Sat follows a very eccentric orbit and only experiences two pericentric passages while D-Sat 
describes a series of fast-decaying orbits and completes many pericentric passages.
At accretion, S-Sat and D-sat
are characterized by similar relative velocities with respect to the host although the impact parameter of D-Sat is three times larger than for S-Sat.

In the central panels, we show the time evolution of the satellite mass in DM (green), gas (blue) and stars (red).  
At accretion, S-Sat has a (total) mass of $\Mac = 2.9 \times 10^8 \, \hMsol$ which progressively reduces to 
$M_{\rm sh} (z=0) = 8.7 \times 10^7 \, \hMsol$. D-Sat is initially much more massive, with
$\Mac = 1.3 \times 10^9 \, \hMsol$. 
The stellar and DM masses remain fairly constant until late times when the satellites reach the densest regions
of the host haloes and are tidally stripped.
On the other hand, the gas component follows a very different evolutionary path.
Even before $\zac$, gas is stripped off the (yet to be) satellites.
This is due to the combined action of two mechanisms:
heating caused by the ultraviolet background radiation and interactions with an increasing environment density.
After the reionization of the intergalactic medium is completed at $z\sim 6$,
photo-heating affects low-density gas which is not self-shielded. As a consequence,
the gas reservoir of haloes that are below the atomic cooling mass limit \citep{Doroshkevich+1967, Rees+1986}
can be completely depleted
\citep{Gnedin2000,Okamoto+2008}.
Simultaneously, as the low-mass haloes approach their final more-massive host, they 
find themselves in denser and denser environments of the cosmic web and can loose their gas via ram pressure stripping
\citet{Benitez-Llambay+2016}. 
Both S-Sat and D-Sat are massive enough to retain a substantial amount of gas at $\zac$. However, 
after they accreted on to their hosts, they are completely deprived of gas within the first few pericentric passages
\citep[see also][]{Helmi+White1999}. 

A more detailed view of the fate of a substructure is provided
by the radial phase space plots in the bottom panels of 
Fig. \ref{fig:phase_space_evolution}.  
Here we show the location of the simulation particles (DM, gas and stars) that form S-Sat and D-Sat
from $z=3.5$ to $z=0$ with time steps of $500 \, \rm{Myr}$. The particles are colour-coded 
based on redshift and, to facilitate understanding, symbols of the corresponding colours are also shown in the top panel.
The effect of tidal interactions is clearly noticeable in the bottom-left panel.
Particles start being stripped off S-Sat during its first pericentric passage.
The debris form tidal tails that eventually fall on to the host along
a very extended stream.
At $z=0$, the original subhalo has been split in
three different components: the material stripped during the first
orbit lying inside the halo; the recently-disrupted material, as
distant as $400 \, \hkpc$, being slowly accreted by the main halo, and the
surviving subhalo orbiting around the host. 
Each pericentric passage ends up generating a dynamically coherent structure in which positions and velocities are strongly correlated. One of these 
`tidal caustics' \citep[e.g.][]{Mohayaee+Shandarin-2006} is clearly noticeable in
the radial phase-space diagram for S-Sat
as a symmetric distribution of loose particles with a uniform colour and a characteristic bell shape.
Note that tidal caustics have a finite width and density 
(contrary to genuine caustics that are generated by a perfectly cold distribution 
of particles with the same energy).
The build up of tidal streams and caustics is particularly evident in the bottom-right panel. Due to its larger mass, D-Sat experiences stronger dynamical
friction than S-Sat. Consequently, it orbits much closer to the centre of the host and with a shorter period, resulting in a rapid disruption. A series of tidal caustics corresponding to the multiple orbits of the satellite are in fact
noticeable in the phase-space diagram of the loose material at $z=0$.

Fig. \ref{fig:phase_space_evolution} shows that, 
after its first pericentric passage, S-Sat reaches a distance
of $\sim 415\ \hkpc$ from its host halo while some of the stripped debris
stretch out to more than $500\ \hkpc$ of separation.
Note that S-Sat travels across its first apocentre at $z_{\rm ap}=0.8$ when the radius of Supay is $\Rh=160\ \hkpc$ (dashed line in the bottom-left panel).
This phenomenon is quite common \citep[see also][and Fig. 8 in Paper I]{Ludlow+2009} and makes it interesting to compare the apocentric distance of the satellite 
with the splashback radius of the host halo which is generally assumed to
mark its physical outer boundary.
We first estimate $\Rspl$ by locating
the minimum of the logarithmic derivative of the spherically averaged mass density profile
\citep{More+2015}. Our result at $z_{\rm ap}$ is shown with a dot-dashed line.
Since numerical derivatives are noisy,
we also make use of the
{\sc shellfish} code \citep{Mansfield+2016} to 
reconstruct the full three-dimensional shape of the splashback surface and derive
$\Rspl$ as the radius of the sphere with the same enclosed volume. 
In all cases, this estimate differs by less than 4 per cent from the previous one.
Finally, we obtain a third value for $\Rspl$ by using a fit to the median splashback radius of haloes with a given mass and accretion rate \citep[averaged over time scales comparable with the halo crossing time, see equation (5) in ][]{More+2015}
and providing the appropriate input for S-Sat. The resulting $\Rspl$ (dotted line)
is always larger than our previous estimates but still substantially lower than the apocentric distance of S-Sat. 
Although we have presented in detail only one specific example, the
same conclusions can be reached after studying the evolution of a very large number of satellites. We find that the first apocentric distance lies beyond the splashback radius for nearly 40 per cent of the satellites, regardless of the host halo. It is worth stressing that two distinct concepts have been mixed up under the name of splashback radius
\citep{Adhikari+2014, More+2015}.
In practical calculations, $\Rspl$ is defined as 
the location at which the halo mass-density profile presents a sudden steepening of its slope. On the other hand, based on spherical models of collisionless secondary infall,
$\Rspl$ is interpreted as the position of the outermost density caustic and, by extension, as the radius at which newly accreted matter reaches its first apocentre.
Our results indicate that, in realistic cases, these two concepts do not perfectly match and the practical definition of $\Rspl$ needs to be further refined in order to make them compatible.
A step forward in this direction has been made by \citet{Diemer2017} who used N-body simulations to investigate the 
relation between $\Rspl$ and the first apocentric distances of DM particles that are not part of substructures. Our complementary study, instead, 
follows the orbits of satellite galaxies in haloes of lower mass.
Further understanding is also required to optimise the strategy for the observational detection of the splashback radius in galaxy clusters using their member galaxies \citep{More+2016,Baxter+2017}.

\subsection{Shed material}
We now focus on the material that was originally locked in substructures and later became part of the host halo and its central galaxy (this study completes the analysis presented in \paperII).
For example, in Fig. \ref{fig:phase_space_z0}, we consider the 100 most massive substructures
identified at $z=2$ in \amun (top) and \supay (bottom) and plot the 
radial phase space distribution of their original constituents (gas, stars and DM) at $z=0$ (similar conclusions can be drawn selecting \abu and \siris, as well as halving or doubling the sample of substructures).
Particles are 
colour-coded according to their temperature (for the gas) or velocity
dispersion (for stars and DM) computed using the 64 nearest neighbours of the same
species. The inset on the bottom of each panel shows the
cumulative probability distribution $P(>r)$ of finding a particle of the given
species at distances larger than $r$.
The phase-space distribution of the gas extends to larger radii with respect to the bulk of the collisionless components.
While stars and DM show clumps associated with surviving satellites, 
virtually all the gas has been stripped off of the substructures. 
Shock-heating and feedback mechanisms have ejected a good fraction of this gas outside the main halo
from where it can rain back in once it has cooled down. However, the cooling time can be quite long depending on the local metallicity and density. 
It is worth stressing that only half of the gas that was part of substructures at $z=2$
is still found within $\Rh$ at $z=0$ (see also the discussion about recycled material in \paperII).
This effect might not be accurately captured by certain `semi-analytic' models of galaxy formation 
that make too simplistic assumptions concerning gas stripping from satellites and stellar feedback
\citep[see also][]{Hirschmann+2012}.

Baryonic material that was at first part of substructures also migrates to the central galaxy and its disc. 
By approximating the latter as a cylinder with height and radius corresponding to three times
the scale values derived in \paperII, we determine the fraction $\fdisc$ 
of the original baryonic mass of a $z=2$ substructure that is found in the galaxy at $z=0$.
This quantity increases with $\Msh$ since
massive subhaloes are more affected by dynamical friction and retain a larger gas reservoir
until disruption. Although $\fdisc$ shows a lot of scatter,
its mean value for each of the ZOMG haloes is rather small, ranging between 3 and 9 per cent
(see Table \ref{table:Halo}). 
The halo assembly time does not seem to have an influence on $\langle \fdisc\rangle$.  
However, this quantity is substantially larger for Amun and Supay that host central galaxies with a prominent disc component 
(see \paperII). 
This is not only reflecting the larger disc size  
but is also a consequence of the deeper potential well that alters the orbits of the satellites
\citep{Garrison-Kimmel+2017}.
Consistently,
the (baryonic) mass fraction of the present-day disc that was inside substructures at $z=2$
amounts to 32 and 8 per cent for Supay and Amun, respectively, while is smaller for Siris (3 per cent) and in particular for Abu ($0.5$ per cent).
Note that only a few massive substructures with large $\fdisc$ determine the value
for Supay.

\section{Subhalo statistics}
\label{sec:Comparison}

After having described the phenomenology and the main physical mechanisms
of satellite evolution, we investigate the statistical properties of the 
surviving substructures.
The goal of this section is twofold. First, we show that
our simulations are in agreement with 
many
observations and previous numerical studies. 
At the same time, however, we try to isolate possible distinctive features
characterizing \old and \young haloes.
Since observation of satellites are only possible in the local universe,
we exclusively present results at $z=0$ (unless explicitly stated otherwise).

\subsection{Mass and radial distribution}

\begin{figure}

\begin{center}
\includegraphics[width=\columnwidth]{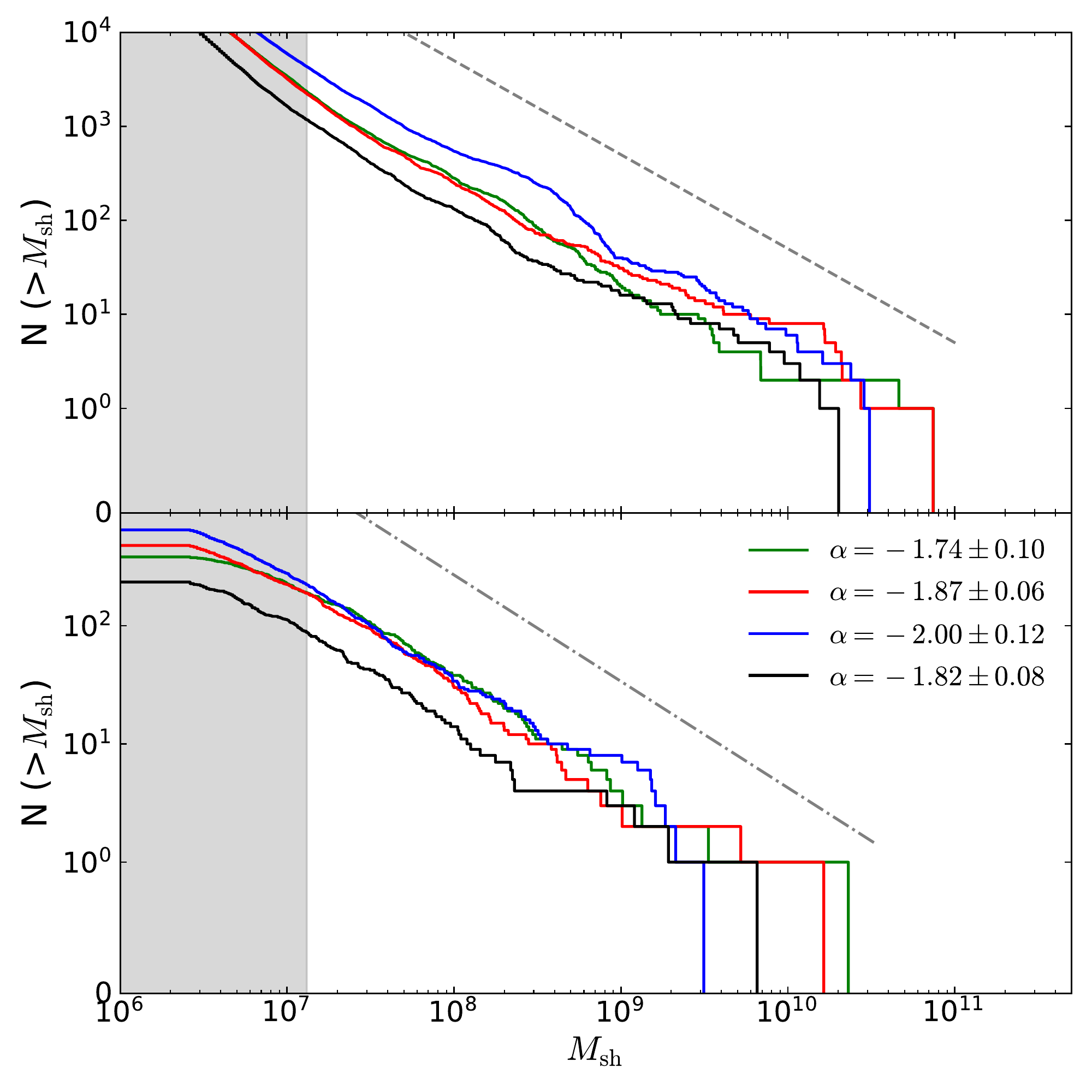} 
\end{center}
\caption{
  Unvolved (top) and evolved (bottom) cumulative subhalo mass functions. The gray 
  vertical shaded region corresponds to subhalo masses smaller than 100 DM particles. 
  The dashed and dot-dashed lines are used as a reference and correspond to a power law with 
  slope $\beta=-1$ and $\alpha+1 = -0.9$, respectively. 
  In the bottom panel the slope of the best-fitting power law for each host are reported.}
\label{fig:sHMF_combo}
\end{figure}

\begin{figure}
\begin{center}
\includegraphics[width=\columnwidth]{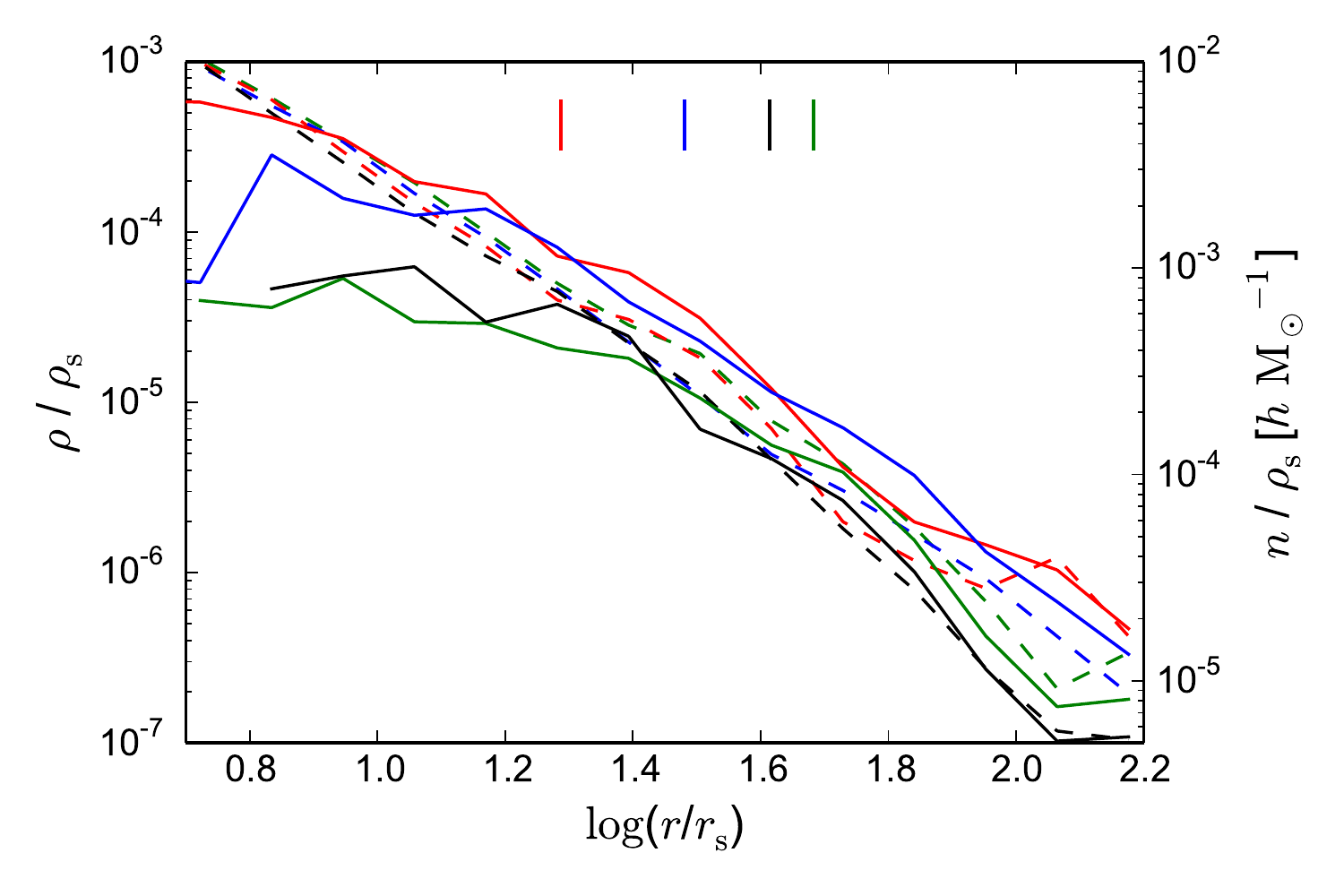}
\end{center}
\caption{Radial distributions of substructures (solid) and matter (dashed) within and around the ZOMG haloes at $z=0$. 
  Note the different vertical scale for the two profiles, which are rescaled by the best-fitting value of $\rho_{\rm s}$.
  Similarly, $r$, the distance from the host 
  centre, is normalized by the scale radius $\rs$. 
  Vertical segments indicates the radius of each halo.}
\label{fig:radial_density}
\end{figure}
 
In the bottom panel of Fig. \ref{fig:sHMF_combo} we show the distribution by mass of satellites
in each of the ZOMG haloes (within $\Rh$ and at $z=0$, commonly dubbed subhalo 
mass function, or sHMF) together with
the slope of
the corresponding best-fitting power law of the form
\begin{equation}
\frac{\d N(\Msh)}{\d \Msh} \propto \Msh^{\alpha}
\end{equation}
(note that, while we show the cumulative sHMF, we perform the fit using the differential distribution of $\Msh$ in order to avoid highly correlated errors).
In the fit, we weigh the binned counts according to their Poisson errors and, 
to limit incompleteness due to resolution effects, we only consider satellites containing at least 100 DM particles (this mass limit is indicated in Fig. \ref{fig:sHMF_combo} with a vertical shaded region).
The best-fitting slopes for the different ZOMG haloes are compatible within the errorbars (the actual values are reported in
Fig. \ref{fig:sHMF_combo}).
Overall, they are in good agreement with the value of $\alpha \simeq -1.9$ generally found in high-resolution $N$-body 
\citep{Klypin+1999,Moore+1999,Ghigna+2000,Aquarius,Gao+2012,Cautun+2014,Coco} and hydrodynamical
\citep{Romano-Diaz+2010, Sawala+2013, Sawala+2017} simulations (dot-dashed line).
This is also consistent with observational estimates based on strong \citep[][for the substructure in galaxy-sized haloes]{Vegetti+2014}
and weak \citep[][for the Coma cluster]{Okabe+2014} gravitational lensing.

In the top panel of Fig. \ref{fig:sHMF_combo}, we
show  the probability
distribution of the satellite masses at accretion for each ZOMG halo.
This unevolved sHMF is usually averaged
over all main haloes present at a given time and includes the satellites 
accreted at any previous time. Numerical studies found it can be well approximated by a power-law
\begin{equation}
\frac{\d N(>\Mac)}{\d \Msh} \propto \Msh^\beta
\end{equation}
 with slope $\beta = -1$ \citep{Giocoli+2008, Aquarius, Han+2016}. Such relation (dashed line) is in good agreement with our results for single haloes.

Within $\Rh$,
the spherically averaged mass density profiles of the ZOMG haloes are well described by the
NFW function
\begin{equation}
\rho(r)= \frac{4\rho_{\rm s}}{(r/\rs) \left[ 1 + (r/\rs)^2 \right]}
\label{eq:NFW}
\end{equation}
where $\rs$ and $\rho_{\rm s} = \rho(\rs)$ denote a characteristic radius and density.
This is shown by the dashed lines in Fig.~\ref{fig:radial_density} where we have appropriately rescaled the horizontal and vertical axes so that the different curves should coincide if the mass profiles exactly follow equation (\ref{eq:NFW}). The corresponding values of $\Rh$ are indicated by vertical segments (see also Table \ref{table:Halo}).
We now compare the radial distribution of the satellites 
in \young and \old haloes at $z=0$ (solid lines in Fig.~\ref{fig:radial_density}). 
All the hosts show the same pattern: the satellite distribution traces the matter profile for 
$r\gtrsim 0.5\, \Rh$ but flattens out at smaller radii where substructures are more easily disrupted.
Similar findings have been originally reported for cluster-sized haloes \citep{Nagai+2005,Gao+2012} and, later,
on galaxy scales based on dark-matter-only \citep[\eg][]{Han+2016}
and hydrodynamical simulations \citep[\eg][]{Libeskind+2007,Schewtschenko+2011,Zhu+2016}.

\subsection{Gas and stellar content}
\label{sec:gas_stellar_content}

\begin{figure}
\begin{center}
\includegraphics[width=\columnwidth]{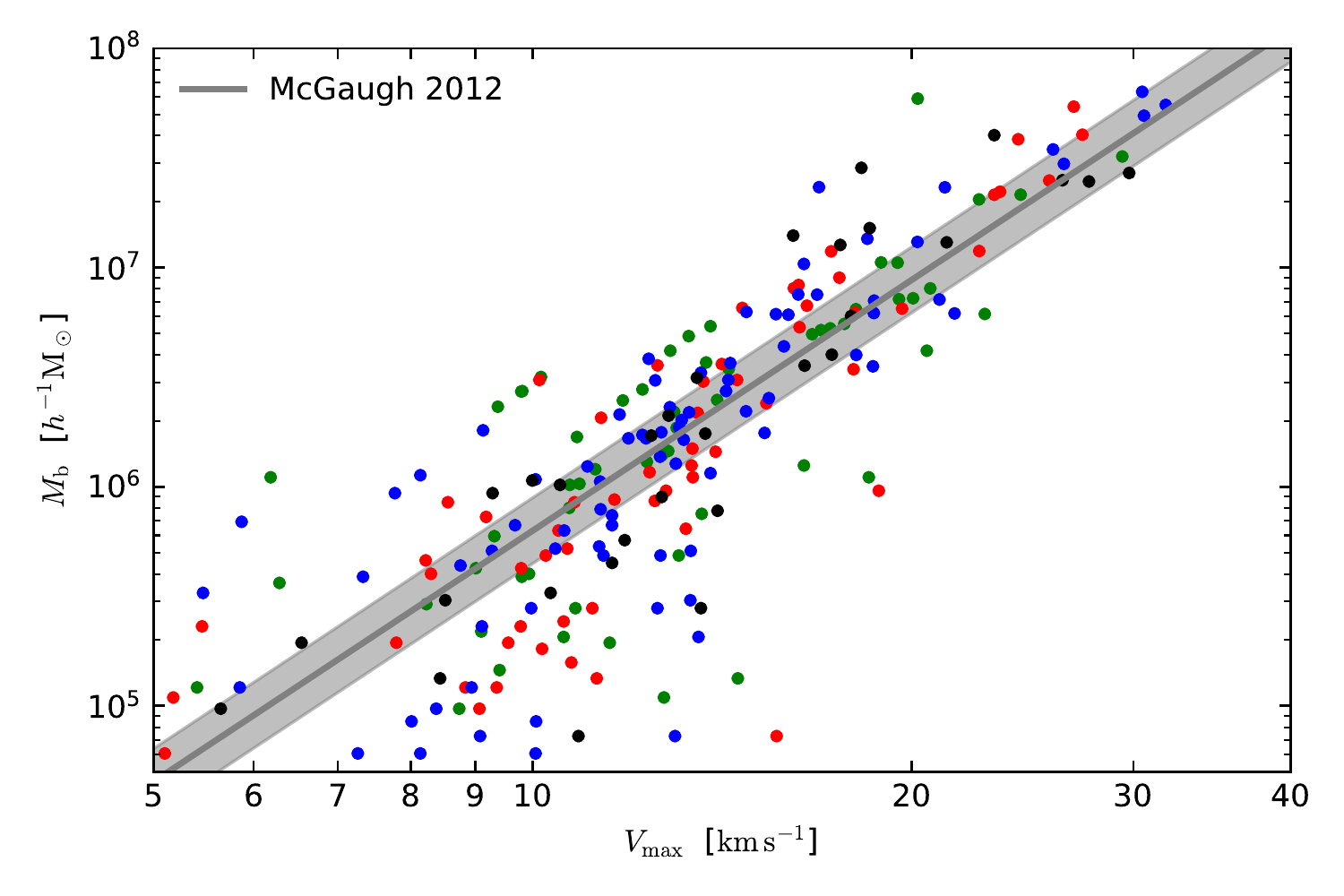}
\end{center}
\caption{Baryonic Tully-Fisher relation for the ZOMG subhalo
  populations. The solid line and the shaded region denote the best-fitting 
  power law and the maximum estimated scatter for the observational data, respectively \protect\citep{McGaugh2012}.}
\label{fig:barTF}
\end{figure}

\begin{figure}
\begin{center}
\includegraphics[width=\columnwidth]{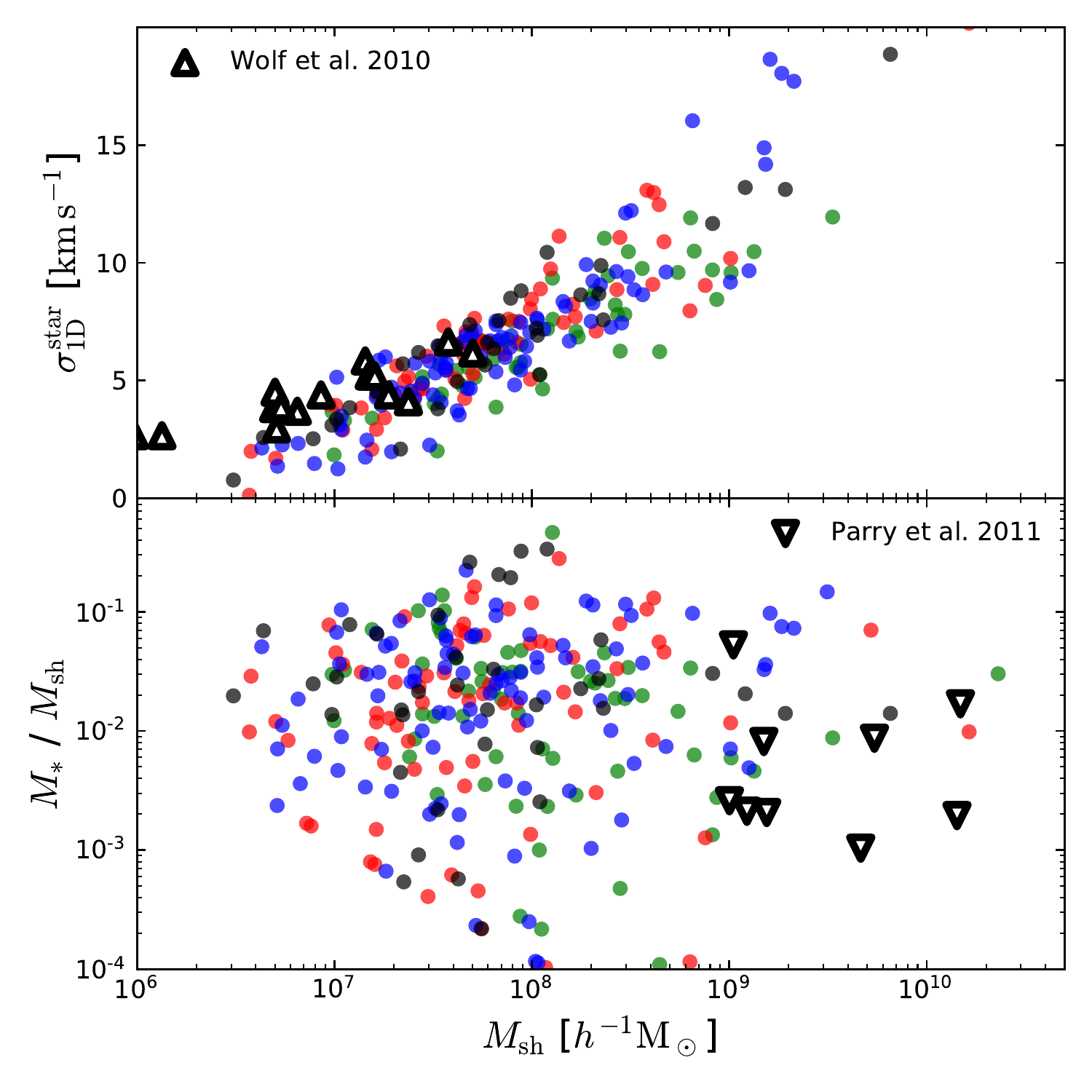}
\end{center}
\caption{Top: 1D velocity dispersion for stellar particles ($\sigma_{\rm
    1D}^{\rm star}$, computed assuming isotropy) as a function of subhalo mass. The
  triangles represent data from \protect\citet{Wolf+2010} for Milky Way
  dwarf spheroidal satellites. Bottom: fraction of mass in stars
  as a function $\Msh$. Black triangles show data from Milky Way
  satellites \protect\citep{Parry+2011}.}
\label{fig:sigma_Mstar_Mtot}
\end{figure}

We now delve into the analysis of the stellar and gas content of
substructures. In Fig. \ref{fig:barTF}, we plot the maximum circular velocity,
$\Vmax$, of a subhalo against 
its total baryonic mass, $M_{\rm b}$.
Observationally, these quantities are tightly correlated and the corresponding power-law relation is known as the baryonic Tully-Fisher (BTF) relation \citep{Tully-Fisher, Walker1999, McGaugh+2000}. 
Since \citet{McGaugh2012} showed that central and satellite galaxies follow consistent relations, 
we compare our data to their fit for both populations (solid gray line).
The simulated substructures are in good agreement with the observed mean relation and no difference is noticeable between the different ZOMG haloes.
It is worth noticing that the simulations overpredict the scatter in the BTF relation, especially at low $\Vmax$. This is a long-known issue with 
the standard scenario of galaxy formation (\eg \citenp{Dutton2012} and \citenp{Lelli+2015}, but see \citenp{Sales+2017} and \citenp{Sorce+2016}).

In Fig. \ref{fig:sigma_Mstar_Mtot} we show scatterplots of the 1D stellar velocity dispersion (top) and the stellar mass fraction (bottom) against
the total mass of the substructures are shown.
The four populations are well mixed and no segregation with $\zc$ is found.
In the common mass range, they match 
data from Milky Way (MW) dwarf spheroidal satellites \citep[][black triangles]{Wolf+2010, Parry+2011}. 

Only a minority of the substructures have a stellar counterpart
(see e.g. the eighth column of Table \ref{table:Halo} and Fig. 
\ref{fig:finder}) which typically formed between redshift $5$ and $6$.
Most satellites, in fact,  
are below the atomic cooling mass limit and are completely 
sterilized during the epoch of reionization. 
Nevertheless, a significant number of massive and concentrated 
subhaloes is able to retain some gas that, sooner or later, contributes to
the gas reservoir of the host halo
(see Fig. \ref{fig:phase_space_z0}) and, possibly, of its central galaxy  (\paperII).
Notably, a relevant portion of the satellites containing some gas 
at $z=0$ have not just accreted on to their hosts (in some cases
$\zac\sim 1$).
Overall, the fraction of substructures with gas in the ZOMG project is consistent
with previous hydrodynamical simulations of cluster-sized haloes
\citep[e.g.][]{Tormen+2004,Dolag+2009}.
Comparing with observations, we note that
the gas content measured in 
Milky Way satellites\footnote{An obvious exception is provided by the Magellanic Clouds.
However, since similar configurations are rare in the \LCDM cosmology \citep[\eg][]{Liu+2011}, it is not surprising 
that we do not find any in our relatively small simulation suite.} \citep{Grcevich+2009, McConnachie2012,
Spekkens+2014, Westmeier+2015} is well below our mass resolution and thus compatible with our gas-naked substructures. 

Finally, we assess the presence of star-forming satellites at $z=0$, defined as substructures containing at least two stellar particles formed in the last $200$ Myr. Interestingly, they are found only
in \young haloes (\amun contains two of them and \abu one). In particular, they are the most massive subhaloes at $z=0$ ($\Msh \gtrsim 5 \times 10^9 \, \hMsol$) although their stellar mass is relatively low ($M_{*, \rm sh} \approx 10^6 \, \hMsol$). They have already experienced one or two pericentric passages and their star-formation rates (SFR) lie in the range $4 \times 10^{-3} < {\rm SFR} \lesssim 1.1 \times 10^{-2} \, \rm M_{\odot} {\rm yr}^{-1}$. Interestingly, a satellite with similar characteristics
(in terms of mass and number of pericentric passages)
is hosted by \supay and did not experience any star formation in the last $200$ Myr. 
At late times, \young haloes 
increase their mass at a faster rate than \old haloes (\paperI). Therefore, they are more likely to host substructures that are massive enough to sustain star formation up to $z=0$. 
We provide further evidence supporting this conjecture in section \ref{sec:fsub}.

\subsubsection{Galactic conformity}
The term `galactic conformity' denotes the tendency of neighbouring galaxies to exhibit
similar colours and SF properties. 
Originally, conformity was detected between galaxies in a single DM halo \citep{Weinmann+2006} although the signal might extend well beyond the virial radius of the host \citep{Kauffmann+2013}.
In particular, it has been shown that passive (star-forming) central galaxies tend to be surrounded by passive (star-forming) satellites.
This trend becomes more and more prominent with decreasing mass of the host halo
\citep[e.g.][]{Weinmann+2006, Knobel+2015, Paranjape+2015}.
In order to extend the theoretical predictions to lower halo masses,
we check if the specific SFR (sSFR, i.e. the SFR per unit stellar mass) of our satellite galaxies reflects the value found for the primary galaxy in each ZOMG halo. 
We follow the strategy of \citet{Weinmann+2006} and classify our galaxies as `quenched' or
`star forming' using a threshold value for the sSFR 
of $10^{-10}\,{\rm yr}^{-1}$. According to this criterion,
all the central and satellite galaxies in our simulations are labelled as quenched at $z=0$, although three of the primaries lie on the star-forming main sequence (see Fig. 17 in \paperII). 
For the satellites, the classification
is independent of the stellar masss, the time interval used to evaluate the SFR and the distance from the central galaxy. In fact, the bulk of the satellites do
not experience any SF in the last few Gyr.
We conclude that the small ZOMG sample shows perfect conformity.

\subsection{Satellite mass fraction}
\label{sec:fsub}

\begin{figure}
\begin{center}
\includegraphics[width=\columnwidth]{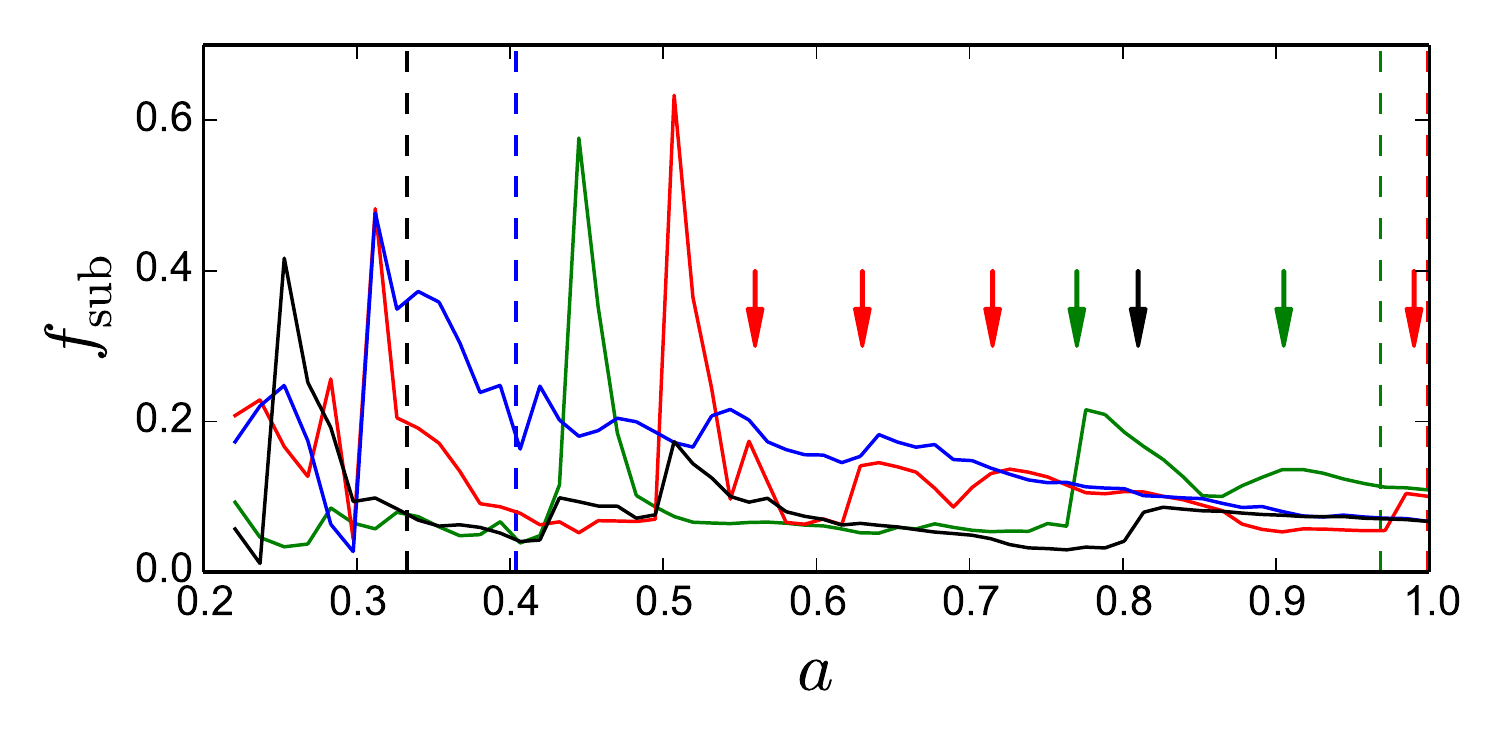}
\end{center}
\caption{Fraction of the host mass contained in resolved substructures ($\fsub$) as a
  function of expansion factor $a$ for the ZOMG haloes. Dotted lines
  denotes the $\ac$ of the host. Arrows highlight when the sharp
  increase in $\fsub$ can be attributed to a massive merger (shown only for $a>0.5$).}
\label{fig:fsub}
\end{figure}

We compute the fraction of $\Mh$ contributed by
(resolved) subhaloes, \ie $\fsub \equiv \sum_i M_{{\rm sh},i} / \Mh$.
At $z=0$, in \young haloes this quantity is 1.6 times larger than in their \old counterparts (see
Table \ref{table:Halo})
and its evolution is shown in Fig. \ref{fig:fsub} as a function of the 
expansion factor. Vertical dashed lines mark $\ac$ for each halo. 
This evolution typically shows many `spikes' where $\fsub$ sharply increases 
and then gradually falls off. 
These peaks correspond to mergers that bring massive satellites inside the main halo 
(indicated by arrows in the figure), sensibly
increasing $\fsub$. Their subsequent drop reflects the mass eroded 
from the satellite that becomes part of the main halo. 
Fig. \ref{fig:fsub} shows that \old haloes have almost no recent ($a<0.5$) 
merger, while \young haloes undergo many merger events in the same period. 
This difference reflects their accretion properties, investigated in \paperI 
and II.

The results illustrated above suggest that \young and \old haloes
host similar subhalo populations at $z=0$ and would be difficult to distinguish observationally. Possible exceptions are: i) the presence of star-forming satellites that only appear in \young haloes, although in small numbers; ii) the different subhalo mass fractions. In the next section, we investigate the assembly history of the subhalo populations and correlate them with the large-scale environments in which \young and \old haloes reside.

\section{Effects of halo assembly}
\label{sec:assembly}

\begin{figure}
\begin{center}
\includegraphics[width=\columnwidth]{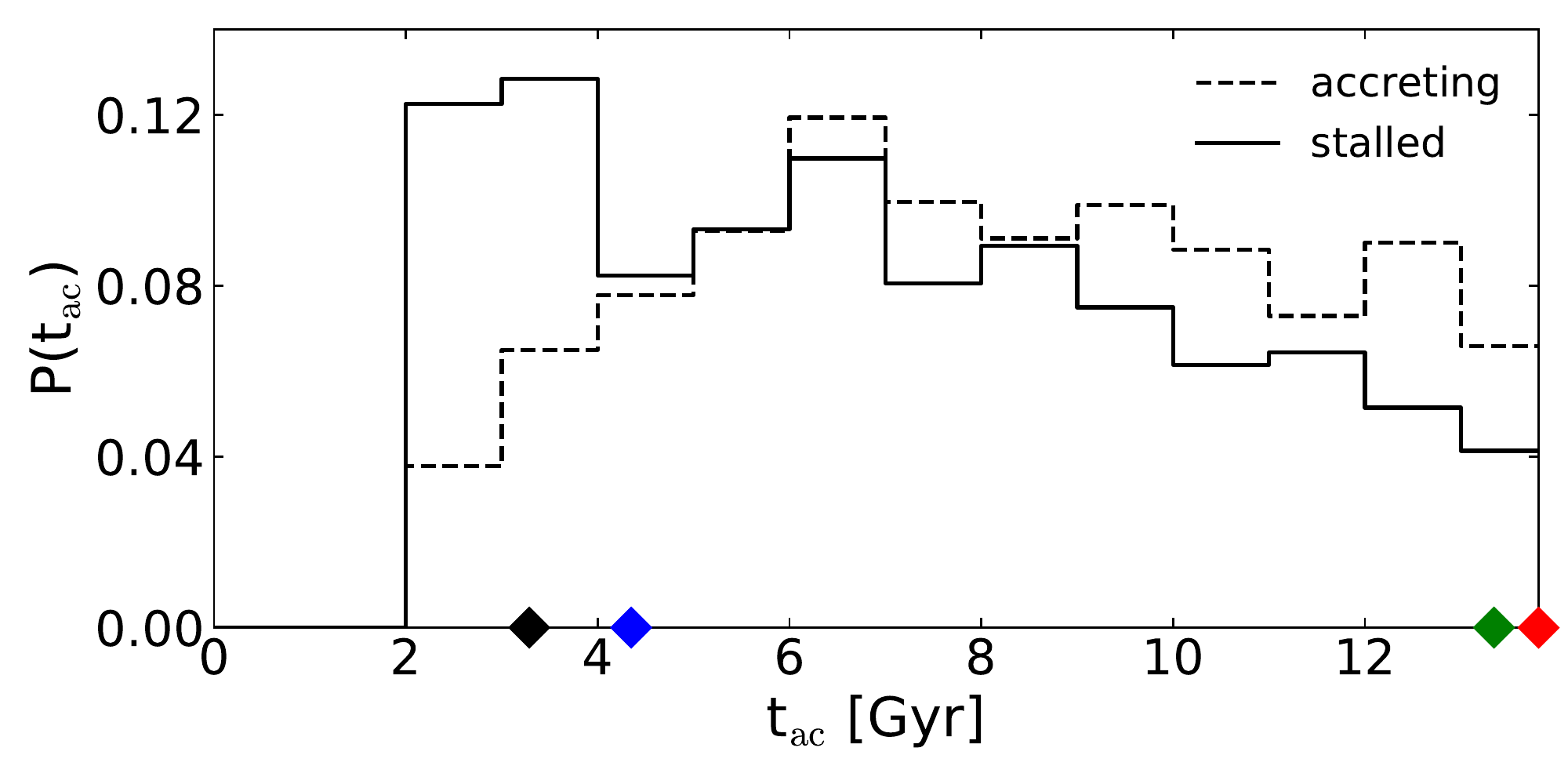}
\end{center}
\caption{Distribution of accretion times of all the substructures
  with $\Msh \geq 100 \, m_{\rm DM}$ that were ever identified within the host. Solid and dashed lines correspond 
  to \old and \young haloes, respectively. Diamond symbols show the assembly
  time of the ZOMG haloes.}
\label{fig:zacc}
\end{figure}

\begin{figure*}
\begin{center}
\includegraphics[width=0.98\textwidth]{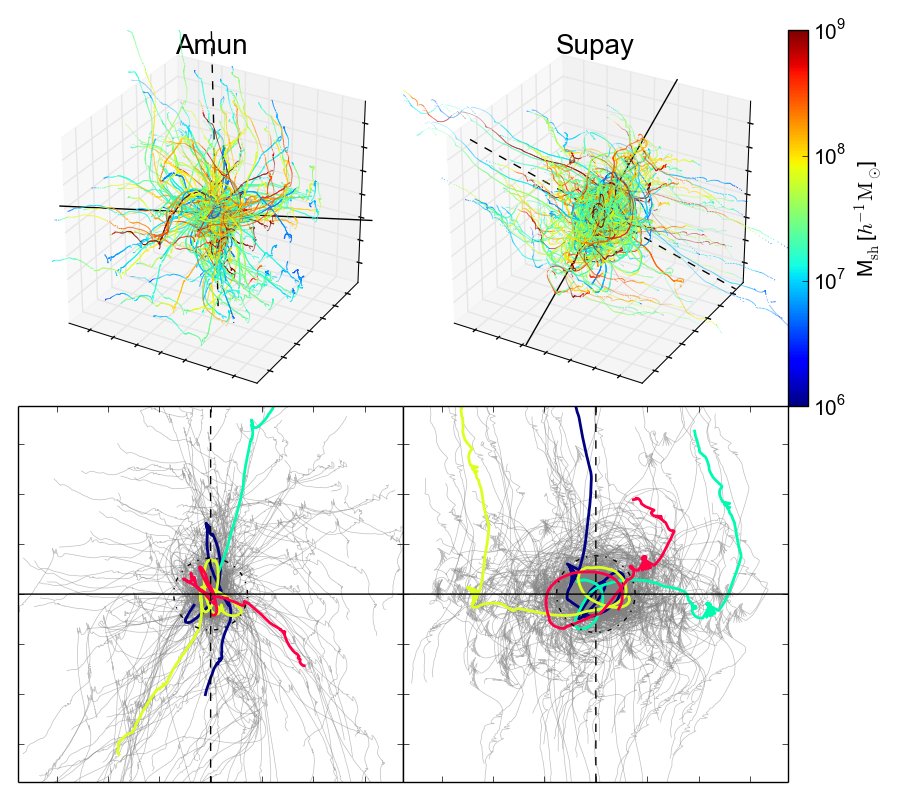}
\end{center}
\caption{Top: Trajectories of all the substructures identified at
  $z=0$ in two haloes of the ZOMG suite. The trajectories are in the host rest
  frame and colour-coded with respect to the subhalo mass. 
  The black solid and dashed lines correspond to the directions of 
  minimum ($\hat{\bf e}_3$) and maximum ($\hat{\bf e}_1$) compression, respectively (see Section
  \ref{sec:AccretionPattern} for details on how they are
  determined). For \supay, $\hat{\bf e}_1$ corresponds to the direction of 
  the filament it is embedded in. Each panel is $1.5 \, \hMpc$ wide in each
  dimension. Bottom: The same as the top panels but projected on
  the plane defined by $\hat{\bf e}_1$ and $\hat{\bf e}_3$. 
  For visual clarity a random subsample of trajectories
  have been highlighted using different colours.
  The dot-dashed circle denotes $\Rh$.}
\label{fig:Trajectories}
\end{figure*}

In \paperI we show that \old haloes reside within prominent filaments of the cosmic web which distort the matter flow
in their surroundings and impact their assembly history and internal dynamics.
On the other hand, \young haloes populate knots of the web and are fed by multiple streams running along thinner filaments.
It is then reasonable to expect that their different accretion patterns might impact the infall of satellite and their final configuration.

\subsection{Accretion time}
\label{accretion_time}

In Fig. \ref{fig:zacc}
we show the distribution of accretion times ($\tac$) for all surviving and disrupted substructures
with $\Msh \geq 100 \, m_{\rm DM}$. 
For comparison, we also mark on the $x$-axis the collapse time of the ZOMG haloes 
using diamond-shaped symbols. 
\Old haloes (solid line) show an enhanced accretion of satellites at very early times ($\tac \leq 4 {\rm Gyr}$) with respect to
\young hosts (dashed line) that present a more prominent tail at late epochs ($\tac \geq 9 {\rm Gyr}$).
The distribution of $\tac$ 
is thus consistent with the description of halo collapse discussed in \paperI.
Interestingly, however, the amount of satellites accreted at late times 
by \young haloes is only marginally larger than in \old hosts.

\subsection{Accretion pattern}
\label{sec:AccretionPattern}

We now investigate if the accretion of satellites follows the
same spatial pattern as the DM. The top panels of Fig. \ref{fig:Trajectories} 
show the trajectories (in the rest frame of the host) of all satellites identified at redshift $z=0$ within
\amun (left panels) and \supay (right panels) -- no significant difference is noticeable in \abu and \siris. 
The colour-coding reflects $\Msh$ as indicated in the bar on the right-hand side.
The black solid and dashed lines highlight the directions of the minor ($\hat{\bf e}_3$) and major ($\hat{\bf e}_1$) axes of inertia for the Lagrangian patch out of which the halo forms (see \paperI for further details).
This region is maximally compressed along $\hat{\bf e}_1$ due to tides. 
For \old haloes, $\hat{\bf e}_3$ instead
coincides with the orientation of the filament they are embedded in.
The bottom panels of Fig. \ref{fig:Trajectories} show the projection of the orbits on the plane 
defined by these two directions. For clarity, a few random trajectories 
are highlighted using coloured lines.
The accretion pattern is different in the two classes of haloes. In \old hosts, substructures first fall in to the
filament along $\hat{\bf e}_1$ and then on to the main halo following curved trajectories.
This secondary infall can only take place within a small region of the filament immediately surrounding the halo.
Beyond this patch, velocities recede from the host along $\hat{\bf e}_3$.
It is exactly this configuration that ultimately suppress matter infall and makes the halo \old. 
We also note that the gravitational field of the filament bends the trajectories of the infalling satellites with respect to 
the radial orbits predicted by many idealized collapse models (\paperI).
On the other hand, \young haloes have a more isotropic accretion pattern 
and their infalling substructures reach the host along approximately radial orbits.
We conclude that the accretion of satellites and DM are regulated by the same dynamics which is manifestly different in \young and \old haloes. Since the distribution and the kinematics of substructures can be constrained with observations, this phenomenon
provides us with a chance to distinguish \young and \old haloes in the local universe.
With this perspective in mind, we explore a number of potential proxies for $\zc$.

\subsection{Subhalo kinematic and anisotropy parameter}
\label{sec:anisotropy}

\begin{figure}
\begin{center}
\includegraphics[width=\columnwidth]{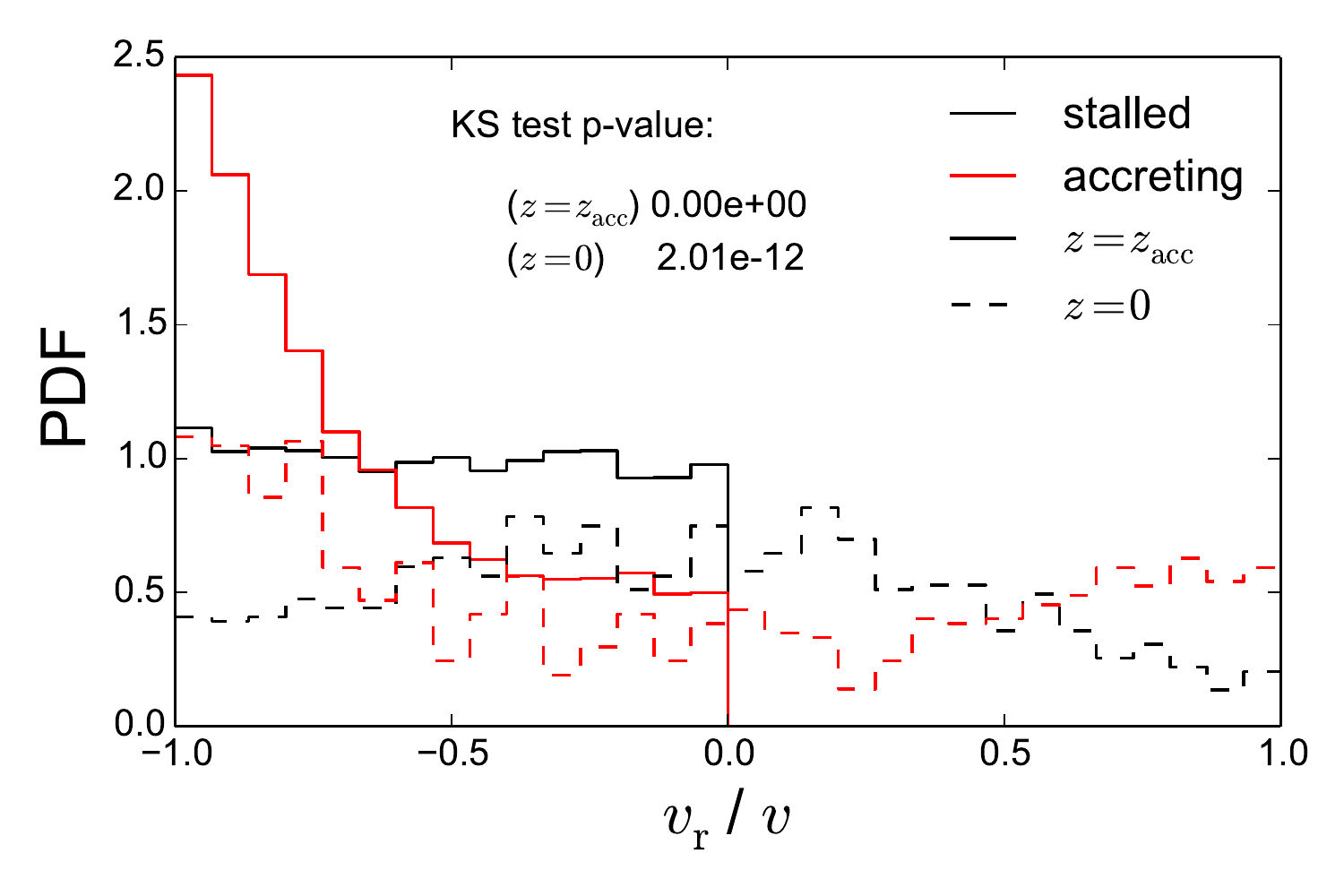}
\end{center}
\caption{Probability density function (PDF) of the (normalized) radial
  velocity at the time of accretion (solid lines) and at $z=0$ (dashed 
  lines) for substructures hosted by \old (black lines) and \young (red lines) haloes.
  Distributions referring to the same time are compared using a Kolmogorov--Smirnov 
  test and the resulting $p$-value is reported.}
\label{fig:vr}
\end{figure}

The velocity orientation of substructures at accretion reflects the different infall
pattern of \young and \old haloes. 
The solid lines in Fig. \ref{fig:vr} show the probability density function (PDF) of the normalized radial velocity component $\vr/\varv$ at accretion for the ZOMG haloes classified based on their collapse time.
In \young haloes (red), the bulk of the satellites accrete on to their hosts along nearly radial orbits. Conversely, the tangential component is more prominent in
\old haloes (black) where the distribution of $\vr / {\varv}$ 
is uniform.
The Kolmogorov--Smirnov (KS) test rules out the null hypothesis that the two samples
of subhalo velocities are extracted from the same population
at a very high confidence level (the corresponding $p$-value is $p<10^{-300}$).

In order to understand if this difference persists with time (and is potentially observable), we analyse the distribution of the satellite velocities at $z=0$
(dashed lines).
The PDFs at $z=0$ are much flatter than at accretion as a consequence of the 
orbit randomization that takes place within the hosts. 
Nevertheless, \young haloes still present an excess of satellites on 
radial orbits (both infalling and outgoing) whilst the substructures of
\old hosts
show a small preference for tangential motion ($\vr / {\varv} \approx 0$).
Even in this case it is very unlikely that the samples are drawn from the same underlying population (the $p$-value of the KS test is $2\times 10^{-12}$).

\begin{figure}
\begin{center}
\includegraphics[width=\columnwidth]{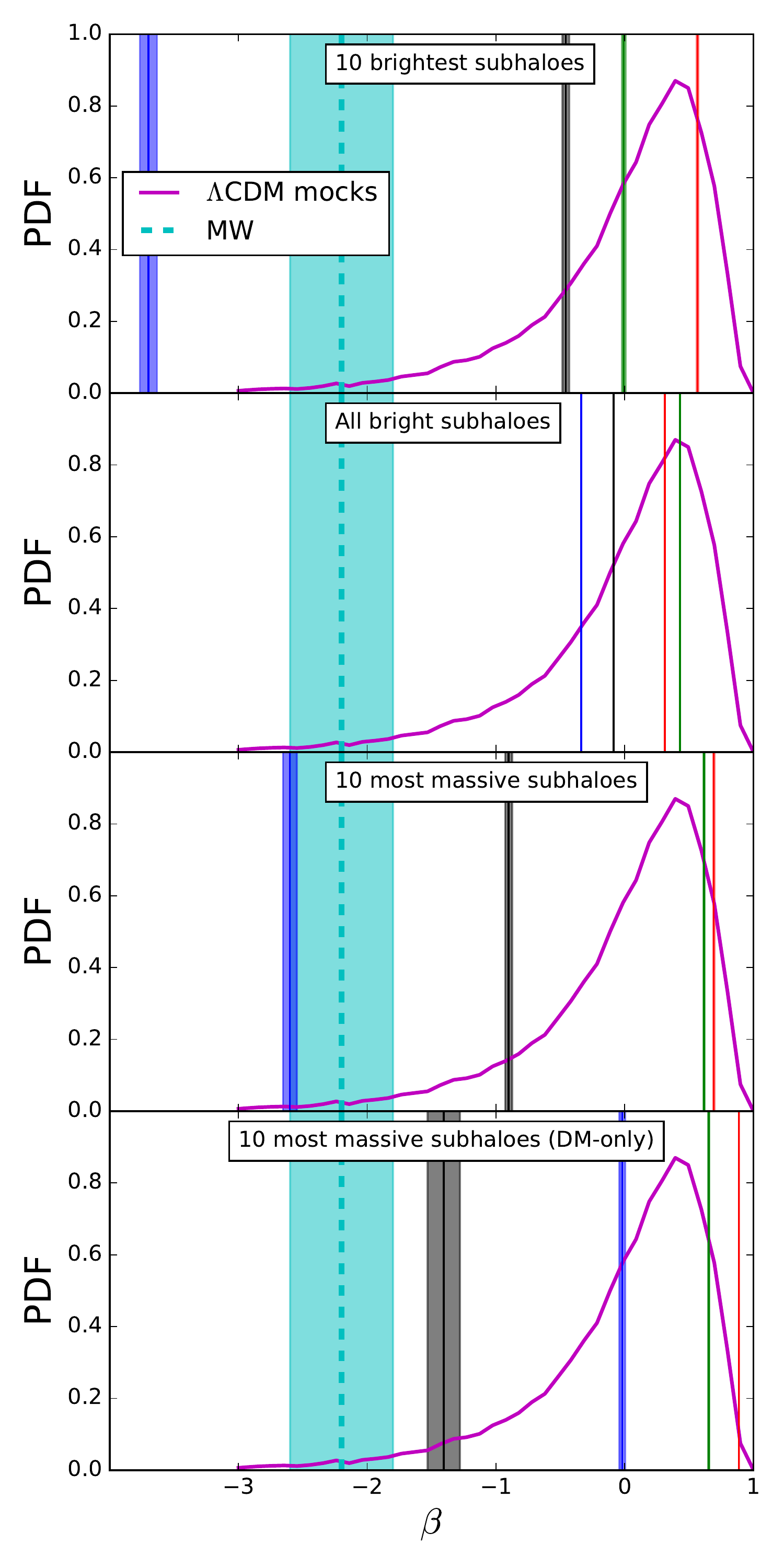}
\end{center}
\caption{Anisotropy parameter $\beta$ computed for our halo sample (solid vertical lines, the shaded region around indicates the bootstrap error). The dashed vertical cyan line (and its shading) shows the results for the MW classical satellites from \protect\citet{Cautun+2016}, while the magenta profile shows the probability distribution function computed by the same authors using a semi-analytical model applied to MW-type DM-haloes. The four panels show different subsets of satellites used to compute the anisotropy parameter, namely (from top to bottom) the ten with largest stellar mass, all bright satellites with at least one stellar particle, the ten most massive in the hydrodynamical runs and in the DM-only ones. 
}
\label{fig:anisotropy}
\end{figure}

From an observational point of view, determining the state of motion of 
a population of satellites around a central galaxy 
is challenging as it requires proper motion measurements.
So far this has been possible only for the closest substructures within the
Milky Way halo but the \textit{Gaia} mission and other future space-based facilities 
might help us extending the measurements to the outer Galaxy and Andromeda 
\citep[e.g.][]{Wilkinson+Evans1999,Kallivayalil+2015}. 
Using 10 satellites with proper motion measurements
(the 11 classical satellites with the exception of Sextans),
\citet{Cautun+2016} conclude that Milky Way satellites show a prominent excess
of tangential orbits which is quite unusual within the $\Lambda$CDM paradigm.
Interestingly, in \paperI, we point out that DM motion in the \old haloes of the ZOMG suite presents enhanced tangential motions triggered by the gravitational effects of 
the filaments they are embedded in.
In order to connect all these findings, we consider 
the anisotropy parameter of the satellites
\begin{equation}
\beta = 1 - \frac{\sum_i \varv^2_{{\rm tan},i}}{2 \sum_i \varv^2_{{\rm rad},i}}
\end{equation}
where $\varv_{{\rm tan},i}$ and $\varv_{{\rm rad},i}$ are the tangential and radial components of the velocity of the $i$-th satellite with respect to the centre
of the host halo and the sum runs over the substructures that fulfill a given selection criterion.
The value of $\beta$ provides a simple parameterization of the satellite
dynamics: $\beta=0$ corresponds to isotropic orbits while positive and negative
values indicate the predominance of tangential and radial orbits, respectively.
In Fig. \ref{fig:anisotropy}, we compare the ZOMG simulations to the results
presented in \citet{Cautun+2016}. The wide vertical shaded band with a dashed line
in the middle indicates the measurement and the uncertainty reported  
for the 10 MW classical satellites. The underlying curve shows the PDF of $\beta$ expected
in the $\Lambda$CDM scenario. This has been extracted
from an $N$-body simulation whose haloes have been populated with galaxies using
a semi-analytical model. To build the PDF,
\citet{Cautun+2016} first select MW-type haloes and then consider
only the 10 satellites with the largest stellar mass for each halo.
Note that only 2.9 per cent of the $\Lambda$CDM haloes are associated with a more extreme value of $\beta$ than the MW.
The values we obtain for the four ZOMG haloes are shown with vertical solid lines surrounded by a shaded band which indicates the corresponding jacknife errors.
To explore the dependence of $\beta$ on the satellite sample we repeat the measurements using four different selection criteria:
the ten satellites with the largest stellar mass (\virg{brightest}, top), all the subhaloes containing at least one stellar particle (\virg{all bright}, upper centre), the ten most massive satellites in the hydrodynamic simulations (\virg{most massive}, lower centre) and in the DM-only runs (bottom). 
It is worth noticing that,
in the hydrodynamic simulations and when we select the 10 most massive subhaloes (either based on the stellar or the total mass), Supay has a more extreme value of $\beta$ than it is measured for the MW.

Although the precise values of $\beta$ vary significantly with the sample, 
\old (\young) haloes invariably show negative (positive) values of $\beta$.
This is a consequence of the large difference in the radial velocities 
at accretion time (see Fig. \ref{fig:vr}), which is preserved until $z=0$. 
Additionally, substructures in \young haloes tend to have larger $\tac$ 
than in \old hosts (see Fig. \ref{fig:zacc}, we check that this holds true
for all the selection criteria employed) that are therefore more affected by orbit randomization.

The results presented in this section suggest a way forward to link observable properties of satellites to the assembly time of their host DM halo. They represent an evidence that satellite dynamics is strongly affected by the environment, which in turn determines the collapse time of the halo (\paperI). Hence, if the proper motion of sufficiently numerous satellites can be measured (\eg in the local universe), \old and \young DM haloes can in principle be distinguished. 
Additionally, in \paperII we show that \old haloes are found to host central galaxies with thicker stellar discs and older stellar populations with respect to \young haloes. 
By considering a large sample of haloes, a dedicated numerical effort could then
determine a precise relation between all these characteristics and $\tc$. 

Based on our results, it is tempting to classify the MW halo as a \old structure.
In fact, the anisotropy parameter of its bright satellites is significantly negative. 
Consistently, the stellar disc of the Galaxy is found to be relatively thick and old \citep{Gilmore+1983, Dalcanton+2002}.
Something to bear in mind is that, contrary to naive expectations, plenty of gas can still accrete on to the central galaxy of a \old halo and sustain regular star formation (see \paperII).

The dependence of the anisotropy parameter on $\tc$ might help shedding new light on another intriguing observational finding. 
Investigating the satellite kinematics in the Sloan Digital Sky Survey, \citet{Wojtak+2013} infer that the satellites of red galaxies tend to have a positive $\beta$. Taken at face value, this result might suggest that red central galaxies are preferentially hosted by \young haloes. In the ZOMG simulations, no obvious correlation has been found 
between the central galaxy type and the halo accretion history (\paperII). However, the ZOMG
suite is too small to detect subtle statistical trends and substantially larger numerical samples are required to draw strong conclusions regarding the interpretation of the observational data.

\subsection{Satellite configuration}
\label{sec:plane}

\begin{table}
\caption{The spatial flatness of a given set of substructures can be determined
by computing the ratio between the lengths of their minor and major axes of inertia ($c/a$). Below we report the values of $c/a$ obtained applying different selection criteria for the satellites. Namely,
MM refers to the ten most massive satellites, MB to the ten brightest, AB to all the substructures with at least one stellar particle, and DM to the ten subhaloes with the largest $\Msh$ in the DM-only run.
We repeat each calculation three times, using different weighting schemes: 
uniform, $w=r^{-1}$ and $w=r^{-2}$, where $r$ denotes the distance of a satellite from the centre of its host halo.}
\begin{tabularx}{\textwidth}{lccc|ccc}

                  															  \\
                  & $w=1$ & $w=r^{-1}$ & $w=r^{-2}$ & $w=1$ & $w=r^{-1}$ & $w=r^{-2}$ \\
\cline{2-7} 
                  &  \multicolumn{3}{c|}{Abu}   &  \multicolumn{3}{c}{Supay} \\
\cline{2-7} 
MM      & 0.09 & 0.19 & 0.33    &    0.52 & 0.61 & 0.33 \\
MB         & 0.17 & 0.26 & 0.32    &    0.14 & 0.18 & 0.20 \\
AB        & 0.43 & 0.49 & 0.53    &    0.44 & 0.46 & 0.44 \\
DM   & 0.12 & 0.20 & 0.33    &    0.12 & 0.19 & 0.22 \\
\cline{2-7} 
                  &  \multicolumn{3}{c|}{Amun}  &  \multicolumn{3}{c}{Siris} \\
\cline{2-7} 
MM      & 0.12 & 0.15 & 0.14    &    0.48 & 0.45 & 0.38 \\
MB         & 0.06 & 0.10 & 0.17    &    0.44 & 0.52 & 0.40 \\
AB        & 0.25 & 0.29 & 0.34    &    0.45 & 0.39 & 0.33 \\
DM   & 0.14 & 0.18 & 0.20    &    0.30 & 0.37 & 0.41 \\
\cline{2-7} 

\end{tabularx}
\label{tab:coa}
\end{table}

\begin{figure}
\begin{center}
\includegraphics[width=\columnwidth]{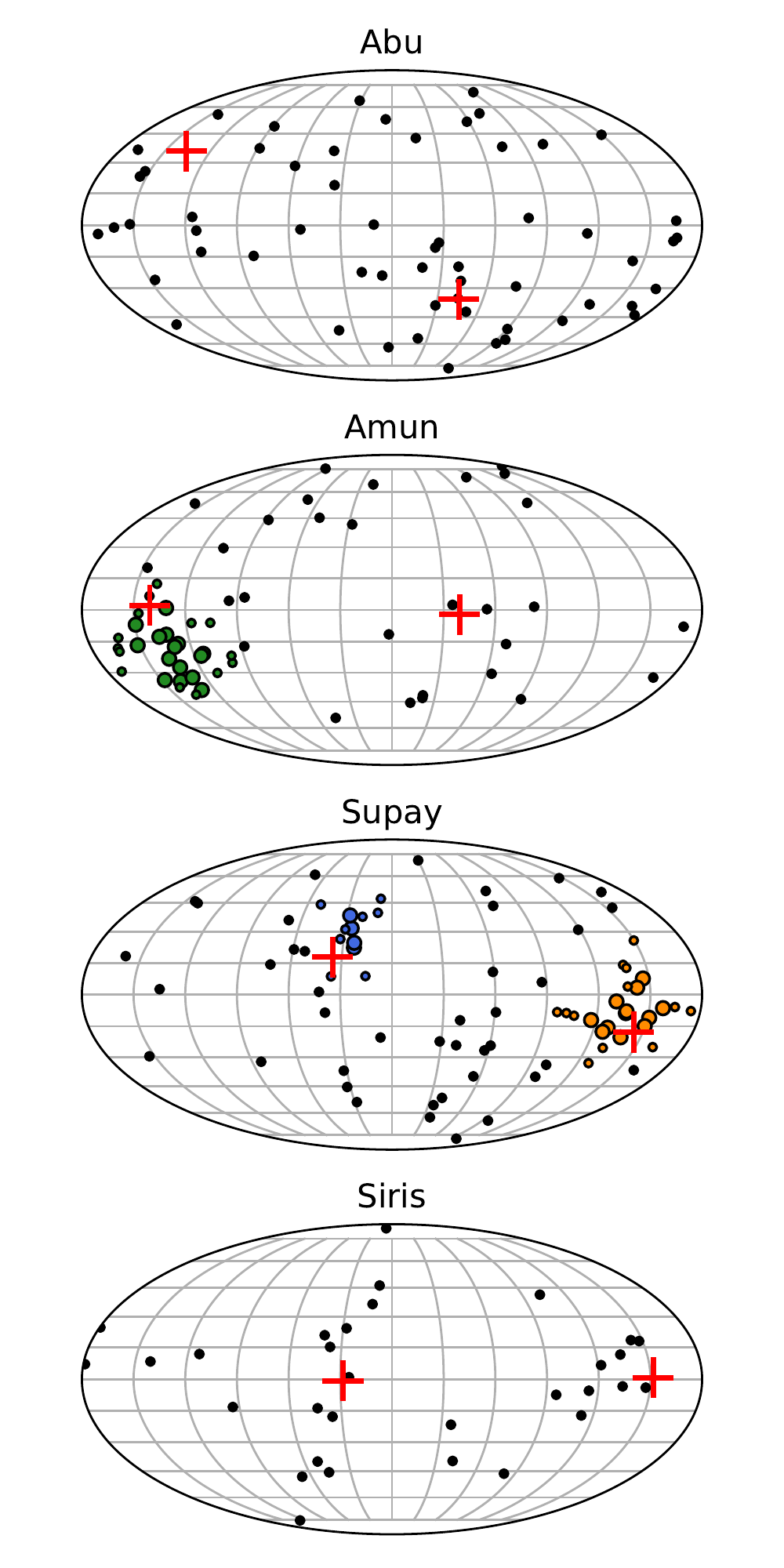}
\end{center}
\caption{Mollweide maps showing the direction of the orbital angular momentum for our simulated satellite galaxies with $M_* > 10^5\,\hMsol$. 
Each \Lcluster identified by DBSCAN is highlighted with a different colour while
black symbols denote unclustered satellites. 
The red crosses indicate the smallest principal axis of the ToI for the `all bright' subsample evaluated with uniform weights.}
\label{fig:clusters}
\end{figure}

\begin{figure*}
\begin{center}
\includegraphics[width=\textwidth]{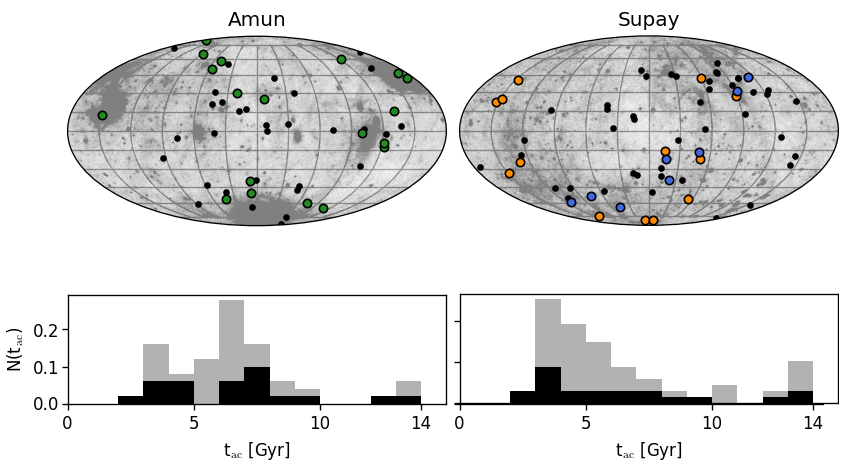}
\end{center}
\caption{Top: Mollweide map of the angular position at accretion time of the satellites with $M_* > 10^5\,\hMsol$ in \supay and \amun (circles, colour-coded as in Fig. \protect\ref{fig:clusters}). In the background, we show the matter distribution at $\Rh$ evaluated at $z=0$ (darker regions correspond to 
higher surface densities). The few thin filaments that converge in \amun are
easily seen whereas the filament embedding \supay is not noticeable because it is
much thicker than $\Rh$.
Bottom: Probability distribution of $\tac$ for all satellites with $M_* > 10^5\,\hMsol$ (gray) and for the satellites in the \Lclusters (black). For \supay both \Lclusters have been combined together.}
\label{fig:accretion_dir}
\end{figure*}

\begin{figure}
\begin{center}
\includegraphics[width=\columnwidth]{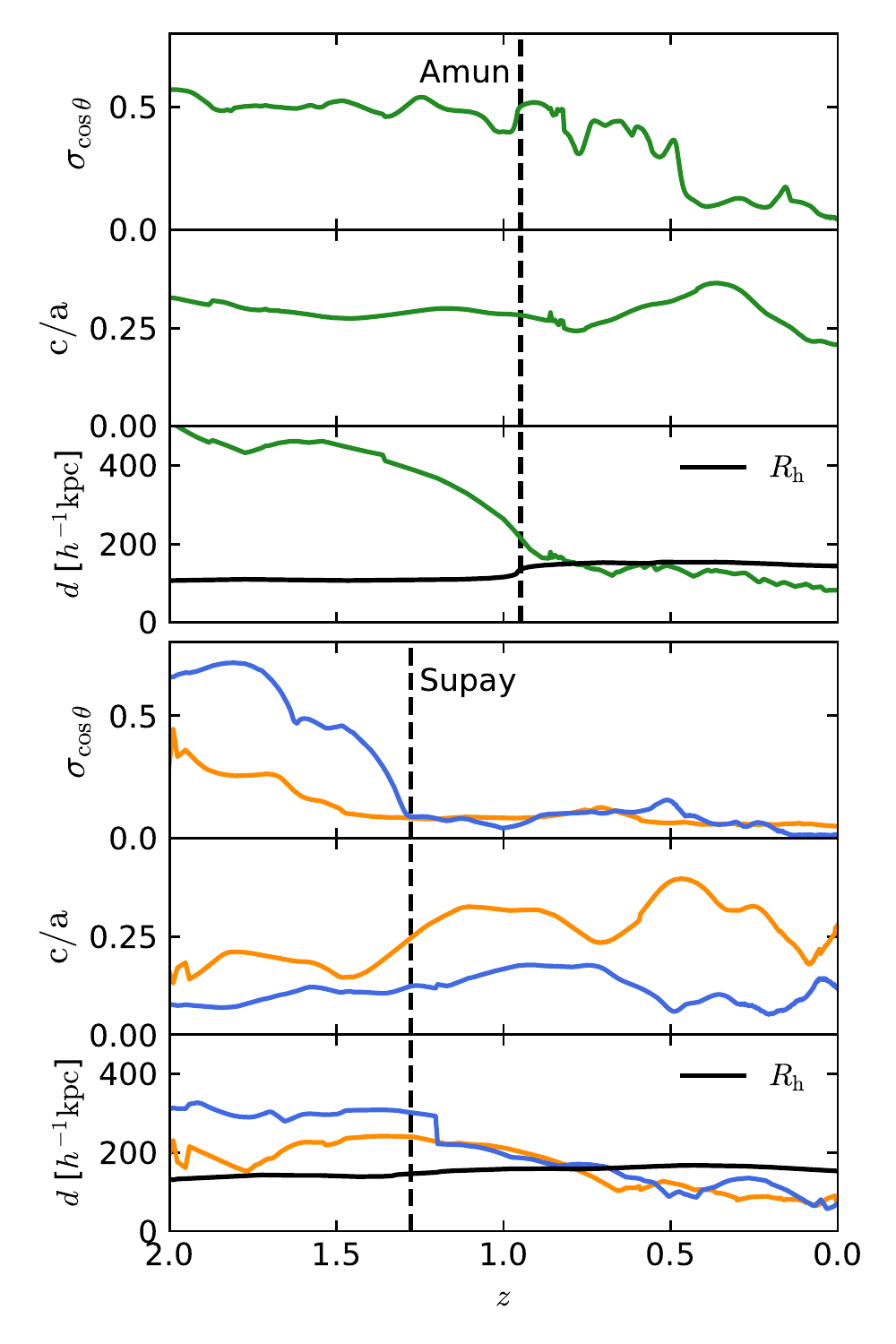}
\end{center}
\caption{The degree of alignment of the orbital angular momenta
($\sigma_{\cos \theta}$, top), the level of planarity 
($c/a$, centre) and the median distance from the host centre ($d$, bottom) are plotted as a function of time
for the satellites that at $z=0$ form the \Lclusters in \supay and \amun. Each \Lcluster is denoted by 
the colour used in Fig. \protect\ref{fig:clusters}. The vertical dashed lines indicate the redshift of the last major merger for the host haloes. For each snapnshot we only consider substructures containing at least 20 DM particles. }
\label{fig:sigma_coa}
\end{figure}

The brightest substructures of the MW lie within a flattened region which is
almost perpendicular to the Galactic disc \citep{Kunkel+Demers1976,Lynden-Bell1976}. 
Proper motion measurements of the classical satellites
suggest that most of them orbit within this `plane' \citep{Kroupa+2005}.
Likewise, nearly half of Andromeda's satellites form 
a thin planar structure \citep{Ibata+2013}.
The existence of similar features has also been reported for
M81 \citep{Chiboucas+2013} and Centaurus A \citep{Tully+2015}.
In addition, the dwarf galaxies within the NGC 3109 association form a filamentary structure which is well ordered in phase space \citep{Bellazzini+2013}.
Within the $\Lambda$CDM paradigm, a few mechanisms have been invoked to explain the presence of these dynamically coherent features. Among these, for instance, are the hypothesis that substructures accrete on to the host while they are part of
clustered groups \citep{Li+Helmi2008, D'Onghia+Lake2008} or that cosmic filaments imprint a preferential direction for
the infall of satellites \citep{Libeskind+2011,Lovell+2011}. 
It is difficult
to conciliate this scenario with the observations in a quantitative way
\citep[e.g.][]{Pawlowski+2012}.
As an alternative, it has been proposed that the planes of satellites in the MW and in Andromeda might originate from the tidal tail of a violent galaxy interaction that took place long ago in the Local Group
\citep[e.g.][and references therein]{Hammer+2013}. Although a consensus has yet to be reached on this issue, it is anyway interesting to further 
explore the implications of the standard cosmological model for the spatial distribution of the satellites. 
For instance, the different accretion modes of \old and \young haloes  
could play a key role in shaping more or less flattened collections of substructures at $z=0$. 
Therefore, we investigate the planarity of the satellite distributions in the ZOMG haloes.

\subsubsection{Spatial distribution}
We first focus on the satellite positions at $z=0$ for the four subsamples we already used in Section \ref{sec:anisotropy}
although a more general analysis will be presented in Section \ref{sec:Lclustering}.
We compute the tensor of inertia (ToI) of the substructures
employing three different weighting schemes (uniform, $r^{-1}$ and $r^{-2}$ where $r$ is the radial distance from the halo centre).
We compute the eigenvalues of the ToI and denote by $a\geq b\geq c$ their square roots.
The degree of planarity of the satellite distribution is quantified using the ratio $c/a$. Our results are presented in Table \ref{tab:coa}. 
Although the measurements sensibly depend on the sample and the weighting scheme,
the brightest (or most massive) satellites in \young haloes tend to lie within flatter spatial configurations with respect to those in \old hosts. 
However, given the variability of the results, it is difficult to draw solid conclusions based on our small sample.
In many cases, the values of $c/a$ extracted from the simulations
are comparable with the estimates\footnote{
\citet{Maji+2017obs, Maji+2017sim} find that estimates of $c/a$ are not stable and argue against the existence of a plane of satellites in the MW.	
\citet{Pawlowski+2017} strongly criticize this claim and emphasize the lack of a proper statistical analysis to reach it.
Our work consists of a theoretical investigation on the impact of the assembly history on the substructure properties.
Drawing conclusions about the MW satellites is outside the scope of this paper and, in any case,
our simulation suite would be too small to conduct such an investigation.
} found for the MW
which range between 0.18 \citep{Metz+2008} and 0.3 \citep{Libeskind+2005}.

We now test whether the flattened satellite distributions in the ZOMG haloes are dynamically stable.
If this is the case, the angular momenta of the satellites should be well aligned (or anti-aligned) with the normal, $\hat{\bf n}$, to the `plane'. 
We therefore compute the angle $\theta_i$ between the orbital angular momentum of a substructure and
the smallest principal axis of the ToI (a proxy for $\hat{\bf n}$).
In many cases,
only a few of the satellites that have been used to identify $\hat{\bf n}$ actually have $|\cos(\theta_i)|\sim 1$.
This suggests that the planar structures determined with the ToI are short lived and probably due to chance alignments. 
In some haloes, however, 60-70 per cent of the selected satellites have $|\cos(\theta_i)|>0.75$ and can orbit in the
flattened structures for longer times. 
This happens, for instance, for both the \young haloes when we consider the most massive substructures
(based on the KS test the difference with the \old haloes is significant at 99 per cent confidence level)
and for \abu and \supay when we select the brightest satellites. Interestingly, $c/a$ is rather low in these cases.

\subsubsection{Clustering of angular momenta}
\label{sec:Lclustering}

To generalize the results presented in the previous section, we now look for groups of substructures with aligned orbital angular
momenta without pre-selecting the satellites based on their structural properties, as we did before to compare
with previous work. 
We only require the stellar mass to be $M_* > 10^5\, \hMsol$ to approximately match the observational limits. 
Coherently rotating groups are identified by applying a clustering algorithm to the directions of the orbital angular momenta. We use the method named
Density-Based Spatial Clustering of Applications with Noise \citep[DBSCAN,][]{DBSCANpaper}  and make sure that our analysis does not depend on its tunable parameters.
The resulting \Lclusters are shown in Fig. \ref{fig:clusters} (coloured filled circles) together with all the unclustered satellites (black circles). The angular momenta of the substructures in \abu and \siris are consistent with a random distribution. 
On the other hand, \amun shows a prominent \Lcluster composed of 25 objects and \supay presents
two of them containing 9 and 21 satellites. Interestingly, these two groups orbit in opposite directions in the plane perpendicular to the vector $\hat{\bf n}$ identified for the `all bright' subsample (red crosses).
It is worth noticing that \amun and \supay are the only two haloes in our simulations that host a grand-design spiral galaxy (\paperII). Hence, it is not inconceivable that related physical mechanisms lie at the origin of both the galactic disc
and the satellite plane. We have checked that the same \Lclusters are also present in the DM-only simulations
suggesting that baryons do not play a key role in their formation process.
Despite the ZOMG sample is only formed of four central galaxies, two of them are surrounded by evident planes of satellites which are dynamically coherent. Our results thus suggest that the occurrence of these structures in the $\Lambda$CDM paradigm is not
unlikely.

In the remainder of this section, we investigate the possible origin of the planar satellite configurations found in 
\amun and \supay. 
In the top panels of Fig. \ref{fig:accretion_dir} we correlate the satellite position at accretion time (circles) with the matter distribution at $\Rh$ (gray background). Substructures are colour-coded as in Fig. \ref{fig:clusters}. Although 
satellites tend to accrete preferentially from high-density regions, the members of each \Lcluster
do not come from the same direction.
Moreover, the distribution of their accretion times fairly traces that of all the luminous satellites (bottom panels in Fig. \ref{fig:accretion_dir}). 
These results do not support filamentary and/or clustered accretion as an explanation for the formation of the planar structures.
 
In order to understand if the \Lclusters are short- or long-lived, we study their time evolution in Fig. \ref{fig:sigma_coa}. We characterize their degree of coherence using the rms
value ($\sigma_{\cos \theta}$) of $\cos \theta$ defined with respect to their mean orbital angular momentum ($\sigma_{\cos \theta}=3^{-1/2}\simeq 0.577$ for a random distribution while $\sigma_{\cos \theta}=0$ for a perfectly coherent cluster).  
Simultaneously, we use $c/a$ (employing uniform weights for the ToI) to describe their flatness and compute the median distance ($d$) of the cluster members from the host.
For both galaxies, $\sigma_{\cos \theta}$ decreases significantly with time while $c/a$ stays approximately constant.
Note that the redshift range covered in Fig. \ref{fig:sigma_coa} extends well beyond the accretion time of the substructures.
As a reference, we indicate the time of the last major merger (determined from the mass accretion history) using a vertical dashed line. 
The \Lclusters in \supay are already identifiable at early times (even for $t<\tc$, at least for the smaller one) when most of the substructures are still beyond
$\Rh$. This is also noticeable in Fig. \ref{fig:accretion_dir} where the angular positions at accretion time of
most cluster members lie within a very flattened region. Based on timing arguments, the formation of the \Lclusters in \supay appears to be connected with the final stages of assembly of the host halo.
On the other hand, the dynamically coherent group in \amun rapidly forms at much later times ($z\sim 0.4$), when more than 50 per cent of its members are already within the host halo. This might possibly reflect the delayed evolution of \young haloes with respect to \old ones.


\section{Conclusions}
\label{sec:conclusions}
We have exploited the high mass and temporal resolution of the ZOMG simulation suite (\paperI, \paperII)
to study the substructure evolution of four haloes with masses of a few$\times 10^{11} \hMsol$ at $z=0$. Abundance matching shows that haloes of this size have most efficiently converted baryons into stars and host $L_*$ galaxies at the present time.
We have characterized the assembly history of the haloes in terms of their collapse redshift, $\zc$, defined as the epoch at which the physical volume enclosing
the halo material first becomes stable.
Galaxy-sized DM haloes identified at $z=0$ show a broad distribution of $\zc$ (see Fig. 2 in \paperI). In \paperI, we have shown that $\zc$ correlates with the cosmic environment surrounding the halo: \old haloes are embedded within prominent filaments of the
cosmic web that inhibit further infall of matter while \young haloes are located
at the knots of the web and are fed by a number of thinner filamentary structures.
In order to study which properties of the haloes (and of the
galaxies within them) depend on $\zc$,
we have selected targets for zoom hydrodynamic simulations by sampling the tails of this distribution. We thus ended up considering
two `\young' haloes ($\zc \lesssim 0$) named \abu and \amun
and two `\old' haloes ($\zc \gtrsim 1$) dubbed \siris and \supay.

Our analysis reveals that many properties of the substructures are
insensitive to the assembly history of the host halo. We list these features below.
\begin{enumerate}  
\item At $z=0$, more than $80$ per cent of the surviving substructures do not contain stars (consistently with the effect of reionization) 
  and more than $99$ per cent are stripped 
  off of the their entire gas content (in agreement with observations of dwarf spheroidal satellites of the Milky Way). 
    
\item Based on our feedback scheme, roughly half of the gas brought in the main halo by satellites is subsequently ejected and remains outside $\Rh$ until $z=0$.    
\item The fraction of baryonic mass in the satellites (identified at $z=2$) that ends up in the disc of the central galaxy at $z=0$ correlates with the disc size
and $\Msh$. This reflects both the strength of the gravitational field generated
by the disc \citep{Garrison-Kimmel+2017} and the increased ability to retain a
gas reservoir that can be then deposited in the central galaxy by the most massive substructures.

\item The first apocentre of nearly 40 per cent of the satellites is
  located beyond the `splashback radius' of the host halo identified as a sudden steepening
  of the mass density profile. This indicates that further work is needed to connect $\Rspl$ with the orbits of recently accreted material. 
   
\item The evolved (i.e. at $z=0$) and unevolved (i.e. at accretion time) mass functions, the radial distribution, the spread of the stellar mass fraction and of the velocity dispersion of the substructures are insensitive to the collapse time of the halo.

\item Based on their sSFR at $z=0$, all the central and satellite galaxies are classified as `quenched' according to the criterion of \citet{Weinmann+2006}.
The ZOMG sample therefore shows perfect conformity between the SF properties of primary and
satellite galaxies and confirms the trend previously found for larger halo masses. 

\item Two of our resimulated haloes (\amun and \supay) contain large clusters of satellites with aligned orbital angular momenta
that form a flattened structure in space. 
The very same features are present in the zoom $N$-body simulations of the 
haloes thus suggesting that baryonic physics does not play a major role in their formation.
Contrary to other studies, we found that
these clusters do not collect satellites that fell in along a specific direction or as a coherent group. 
Intriguingly, \amun and \supay are the only haloes in our sample that host a grand-design spiral galaxy at their centre thus suggesting a possible connection between the physics of disc formation and the assembly of planar configurations of satellites.
\end{enumerate}

On the other hand, additional properties of the substructures clearly depend
on the assembly history of the host halo. Most of them are related to the spatial
and temporal pattern of satellite accretions that directly reflect
the different cosmic environments hosting \young and \old haloes. 
They can be summarized as follows.

\begin{enumerate} \setcounter{enumi}{7}
\item The fraction of halo mass locked in substructures at $z=0$ is substantially larger in \young haloes. Concurrently, \old (\young) haloes accrete a larger fraction of satellites at early (late) times.

\item The epoch at which the planar structures of satellites are formed (see item (vi) above) seems to be connected with the assembly time of the host haloes.

\item Substructures fall in towards \young haloes following nearly radial trajectories.
On the contrary, satellites initially orbit the filament that embeds the \old
haloes before falling on to their hosts. For this reason, they have a large
tangential velocity component at accretion time. 
  
\item Although maximal at infall,
the different balance between the radial and tangential components of the satellite velocities in \young and \old haloes is clearly noticeable also at later times. For instance, the velocity anisotropy parameter of the satellites at $z=0$
is positive for \young haloes and negative for \old haloes. This finding parallels the result found in \paperI   
for the DM particles and provides a tool to determine the formation time of a halo based on the kinematic properties of its
satellite galaxies. 
\citet{Cautun+2016} have recently measured a strong tangential excess for the classical MW satellites
corresponding to an anisotropy parameter of $\beta=-2.2\pm 0.4$. It is thus tempting to tentatively categorize the MW
halo as \old. Further support to this conjecture comes from the thickness and age of the stellar disc
(\paperII).
\end{enumerate}

Future work will aim to extend the approach presented in this work to a larger sample of haloes 
covering a wider range of halo masses and collapse times.

\section*{Acknowledgements}
We thank the anonymous referee for suggesting 
to discuss galactic conformity, 
Marius Cautun for useful discussions,
Philip Mansfield for help with the {\sc shellfish} 
code and Volker Springel for making \pgadget available 
to us. This work is carried out within the SFB 956 
\virg{The Conditions and Impact of Star Formation}, sub-project C4, and 
the Transregio 33 \virg{The Dark Universe} projects funded by the 
Deutsche Forschungsgemeinschaft (DFG). MB thanks 
the Bonn-Cologne Graduate School for Physics and Astronomy for
support. The results presented were achieved employing computing
resources (Cartesius) at SURF/SARA, The Netherlands as part of the
PRACE-3IP project (FP7 RI-312763). We are thankful to the community
developing and maintaining software packages extensively used
in our work, namely: Matplotlib \citep{matplotlib},
NumPy \citep{numpy}, scikit-learn \citep{sklearn}, SciPy \citep{scipy}.


\bibliographystyle{mnras}
\bibliography{substructures}{}

\bsp	
\label{lastpage}
\end{document}